
\input phyzzx
\def\havefigures{n}      
\catcode`\@=12        
%
\newbox\hdbox%
\newcount\hdrows%
\newcount\multispancount%
\newcount\ncase%
\newcount\ncols
\newcount\nrows%
\newcount\nspan%
\newcount\ntemp%
\newdimen\hdsize%
\newdimen\newhdsize%
\newdimen\parasize%
\newdimen\spreadwidth%
\newdimen\thicksize%
\newdimen\thicksz
\newdimen\thinsize%
\newdimen\tablewidth%
\newif\ifcentertables%
\newif\ifendsize%
\newif\iffirstrow%
\newif\iftableinfo%
\newtoks\dbt%
\newtoks\hdtks%
\newtoks\savetks%
\newtoks\tableLETtokens%
\newtoks\tabletokens%
\newtoks\widthspec%
%
%
%
%
\tableinfotrue%
\catcode`\@=11
%
%
\def\tstrut{\vrule height3.1ex depth1.2ex width0pt}%
\def\and{\char`\&}
\def\tablerule{\noalign{\hrule height\thinsize depth0pt}}%
\thicksize=1.5pt
\thinsize=0.6pt
\thicksz=1.5pt
\def\thickrule{\noalign{\hrule height\thicksize depth0pt}}%
\def\ctr#1{\hfil\ #1\hfil}%
%
%
%
%
\tablewidth=-\maxdimen%
\spreadwidth=-\maxdimen%
\def\tabskipglue{0pt plus 1fil minus 1fil}%
%
%
\centertablestrue%
%
%
%
%
\parasize=4in%
\gdef\ARGS{########}
\gdef\headerARGS{####}
\def\@mpersand{&}
{\catcode`\|=13
\gdef\letbarzero{\let|0}
\gdef\letbartab{\def|{&&}}%
\gdef\letvbbar{\let\vb|}%
}
{\catcode`\&=4
\def\ampskip{&\omit\hfil&}
\catcode`\&=13
\let&0
\xdef\letampskip{\def&{\ampskip}}%
\gdef\letnovbamp{\let\novb&\let\tab&}
}
\def\begintable{
   \begingroup%
   \catcode`\|=13\letbartab\letvbbar%
   \catcode`\&=13\letampskip\letnovbamp%
   \def\multispan##1{
      \omit \mscount##1%
      \multiply\mscount\tw@\advance\mscount\m@ne%
      \loop\ifnum\mscount>\@ne \sp@n\repeat%
   }
   \def\|{%
      &\omit\widevline&%
   }%
   \ruledtable
}
\long\def\ruledtable#1\endtable{%
%
%
%
   \offinterlineskip
   \tabskip 0pt
   \def\widevline{\vrule width\thicksz}
   \def\endrow{\@mpersand\omit\hfil\crnorm\@mpersand}%
   \def\crthick{\@mpersand\crnorm\thickrule\@mpersand}%
   \def\crthickneg##1{\@mpersand\crnorm\thickrule
	  \noalign{{\skip0=##1\vskip-\skip0}}\@mpersand}%
   \def\crnorule{\@mpersand\crnorm\@mpersand}%
   \def\crnoruleneg##1{\@mpersand\crnorm
	  \noalign{{\skip0=##1\vskip-\skip0}}\@mpersand}%
   \let\nr=\crnorule
   \def\endtable{\@mpersand\crnorm\thickrule}%
   \let\crnorm=\cr
%
%
   \edef\cr{\@mpersand\crnorm\tablerule\@mpersand}%
   \def\crneg##1{\@mpersand\crnorm\tablerule
	  \noalign{{\skip0=##1\vskip-\skip0}}\@mpersand}%
   \let\ctneg=\crthickneg
   \let\nrneg=\crnoruleneg
   \the\tableLETtokens
%
%
   \tabletokens={&#1}
%
%
   \countROWS\tabletokens\into\nrows%
   \countCOLS\tabletokens\into\ncols%
%
%
   \advance\ncols by -1%
   \divide\ncols by 2%
   \advance\nrows by 1%
%
%
   \iftableinfo %
      \immediate\write16{[Nrows=\the\nrows, Ncols=\the\ncols]}%
   \fi%
%
%
   \ifcentertables
      \ifhmode \par\fi
      \line{
      \hss
   \else %
      \hbox{%
   \fi
      \vbox{%
	 \makePREAMBLE{\the\ncols}
	 \edef\next{\preamble}
	 \let\preamble=\next
	 \makeTABLE{\preamble}{\tabletokens}
      }
      \ifcentertables \hss}\else }\fi
   \endgroup
   \tablewidth=-\maxdimen
   \spreadwidth=-\maxdimen
}
\def\makeTABLE#1#2{
   {
   \let\ifmath0
   \let\header0
   \let\multispan0
%
%
   \ncase=0%
   \ifdim\tablewidth>-\maxdimen \ncase=1\fi%
   \ifdim\spreadwidth>-\maxdimen \ncase=2\fi%
   \relax
%
   \ifcase\ncase %
      \widthspec={}%
   \or %
      \widthspec=\expandafter{\expandafter t\expandafter o%
		 \the\tablewidth}%
   \else %
      \widthspec=\expandafter{\expandafter s\expandafter p\expandafter r%
		 \expandafter e\expandafter a\expandafter d%
		 \the\spreadwidth}%
   \fi %
   \xdef\next{
      \halign\the\widthspec{%
      #1
      \noalign{\hrule height\thicksize depth0pt}
%
      \the#2\endtable
%
      }
   }
   }
   \next
}
\def\makePREAMBLE#1{
   \ncols=#1
   \begingroup
   \let\ARGS=0
   \edef\xtp{\widevline\ARGS\tabskip\tabskipglue%
   &\ctr{\ARGS}\tstrut}
   \advance\ncols by -1
   \loop
      \ifnum\ncols>0 %
      \advance\ncols by -1%
      \edef\xtp{\xtp&\vrule width\thinsize\ARGS&\ctr{\ARGS}}%
   \repeat
   \xdef\preamble{\xtp&\widevline\ARGS\tabskip0pt%
   \crnorm}
   \endgroup
}
\def\countROWS#1\into#2{
   \let\countREGISTER=#2%
   \countREGISTER=0%
   \expandafter\ROWcount\the#1\endcount%
}%
\def\ROWcount{%
   \afterassignment\subROWcount\let\next= %
}%
\def\subROWcount{%
   \ifx\next\endcount %
      \let\next=\relax%
   \else%
      \ncase=0%
      \ifx\next\cr %
	 \global\advance\countREGISTER by 1%
	 \ncase=0%
      \fi%
      \ifx\next\endrow %
	 \global\advance\countREGISTER by 1%
	 \ncase=0%
      \fi%
      \ifx\next\crthick %
	 \global\advance\countREGISTER by 1%
	 \ncase=0%
      \fi%
      \ifx\next\crnorule %
	 \global\advance\countREGISTER by 1%
	 \ncase=0%
      \fi%
      \ifx\next\crthickneg %
	 \global\advance\countREGISTER by 1%
	 \ncase=0%
      \fi%
      \ifx\next\crnoruleneg %
	 \global\advance\countREGISTER by 1%
	 \ncase=0%
      \fi%
      \ifx\next\crneg %
	 \global\advance\countREGISTER by 1%
	 \ncase=0%
      \fi%
      \ifx\next\header %
	 \ncase=1%
      \fi%
      \relax%
      \ifcase\ncase %
	 \let\next\ROWcount%
      \or %
	 \let\next\argROWskip%
      \else %
      \fi%
   \fi%
   \next%
}
\def\counthdROWS#1\into#2{%
\dvr{10}%
   \let\countREGISTER=#2%
   \countREGISTER=0%
\dvr{11}%
\dvr{13}%
   \expandafter\hdROWcount\the#1\endcount%
\dvr{12}%
}%
\def\hdROWcount{%
   \afterassignment\subhdROWcount\let\next= %
}%
\def\subhdROWcount{%
   \ifx\next\endcount %
      \let\next=\relax%
   \else%
      \ncase=0%
      \ifx\next\cr %
	 \global\advance\countREGISTER by 1%
	 \ncase=0%
      \fi%
      \ifx\next\endrow %
	 \global\advance\countREGISTER by 1%
	 \ncase=0%
      \fi%
      \ifx\next\crthick %
	 \global\advance\countREGISTER by 1%
	 \ncase=0%
      \fi%
      \ifx\next\crnorule %
	 \global\advance\countREGISTER by 1%
	 \ncase=0%
      \fi%
      \ifx\next\header %
	 \ncase=1%
      \fi%
\relax%
      \ifcase\ncase %
	 \let\next\hdROWcount%
      \or%
	 \let\next\arghdROWskip%
      \else %
      \fi%
   \fi%
   \next%
}%
{\catcode`\|=13\letbartab
\gdef\countCOLS#1\into#2{%
   \let\countREGISTER=#2%
   \global\countREGISTER=0%
   \global\multispancount=0%
   \global\firstrowtrue
   \expandafter\COLcount\the#1\endcount%
   \global\advance\countREGISTER by 3%
   \global\advance\countREGISTER by -\multispancount
}%
\gdef\COLcount{%
   \afterassignment\subCOLcount\let\next= %
}%
{\catcode`\&=13%
\gdef\subCOLcount{%
   \ifx\next\endcount %
      \let\next=\relax%
   \else%
      \ncase=0%
      \iffirstrow
	 \ifx\next& %
	    \global\advance\countREGISTER by 2%
	    \ncase=0%
	 \fi%
	 \ifx\next\span %
	    \global\advance\countREGISTER by 1%
	    \ncase=0%
	 \fi%
	 \ifx\next| %
	    \global\advance\countREGISTER by 2%
	    \ncase=0%
	 \fi
	 \ifx\next\|
	    \global\advance\countREGISTER by 2%
	    \ncase=0%
	 \fi
	 \ifx\next\multispan
	    \ncase=1%
	    \global\advance\multispancount by 1%
	 \fi
	 \ifx\next\header
	    \ncase=2%
	 \fi
	 \ifx\next\cr	    \global\firstrowfalse \fi
	 \ifx\next\endrow   \global\firstrowfalse \fi
	 \ifx\next\crthick  \global\firstrowfalse \fi
	 \ifx\next\crnorule \global\firstrowfalse \fi
	 \ifx\next\crnoruleneg \global\firstrowfalse \fi
	 \ifx\next\crthickneg  \global\firstrowfalse \fi
	 \ifx\next\crneg       \global\firstrowfalse \fi
      \fi
\relax
      \ifcase\ncase %
	 \let\next\COLcount%
      \or %
	 \let\next\spancount%
      \or %
	 \let\next\argCOLskip%
      \else %
      \fi %
   \fi%
   \next%
}%
\gdef\argROWskip#1{%
   \let\next\ROWcount \next%
}
\gdef\arghdROWskip#1{%
   \let\next\ROWcount \next%
}
\gdef\argCOLskip#1{%
   \let\next\COLcount \next%
}
}
}
\def\spancount#1{
   \nspan=#1\multiply\nspan by 2\advance\nspan by -1%
   \global\advance \countREGISTER by \nspan
   \let\next\COLcount \next}%
\def\dvr#1{\relax}%
\def\header#1{%
\dvr{1}{\let\cr=\@mpersand%
\hdtks={#1}%
\counthdROWS\hdtks\into\hdrows%
\advance\hdrows by 1%
\ifnum\hdrows=0 \hdrows=1 \fi%
\dvr{5}\makehdPREAMBLE{\the\hdrows}%
\dvr{6}\getHDdimen{#1}%
{\parindent=0pt\hsize=\hdsize{\let\ifmath0%
\xdef\next{\valign{\headerpreamble #1\crnorm}}}\dvr{7}\next\dvr{8}%
}%
}\dvr{2}}
\def\makehdPREAMBLE#1{
\dvr{3}%
\hdrows=#1
{
\let\headerARGS=0%
\let\cr=\crnorm%
\edef\xtp{\vfil\hfil\hbox{\headerARGS}\hfil\vfil}%
\advance\hdrows by -1
\loop
\ifnum\hdrows>0%
\advance\hdrows by -1%
\edef\xtp{\xtp&\vfil\hfil\hbox{\headerARGS}\hfil\vfil}%
\repeat%
\xdef\headerpreamble{\xtp\crcr}%
}
\dvr{4}}
\def\getHDdimen#1{%
\hdsize=0pt%
\getsize#1\cr\end\cr%
}
\def\getsize#1\cr{%
\endsizefalse\savetks={#1}%
\expandafter\lookend\the\savetks\cr%
\relax \ifendsize \let\next\relax \else%
\setbox\hdbox=\hbox{#1}\newhdsize=1.0\wd\hdbox%
\ifdim\newhdsize>\hdsize \hdsize=\newhdsize \fi%
\let\next\getsize \fi%
\next%
}%
\def\lookend{\afterassignment\sublookend\let\looknext= }%
\def\sublookend{\relax%
\ifx\looknext\cr %
\let\looknext\relax \else %
   \relax
   \ifx\looknext\end \global\endsizetrue \fi%
   \let\looknext=\lookend%
    \fi \looknext%
}%
%
%
\def\tablelet#1{%
   \tableLETtokens=\expandafter{\the\tableLETtokens #1}%
}%
\catcode`\@=12
%


\font\twelvembf=cmmib10 scaled\magstep1

\font\eighti=cmmi8                          \skewchar\eighti='177
\font\eightsy=cmsy8                          \skewchar\eightsy='60
\font\eightsl=cmsl8
\font\eightit=cmti8
\def\noblackbox{\overfullrule=0pt}
\noblackbox
\def\half{{1\over2}}

\def\bold#1{\setbox0=\hbox{$#1$}%
     \kern-.025em\copy0\kern-\wd0
     \kern.05em\copy0\kern-\wd0
     \kern-.025em\raise.0433em\box0 }
\def\unlock{\catcode`@=11} 
\def\lock{\catcode`@=12} 
\def\Buildrel#1\under#2{\mathrel{\mathop{#2}\limits_{#1}}}
\def\llongrarrow{\hbox to 40pt{\rightarrowfill}}

%
 \newtoks\slashfraction
 \slashfraction={.13}
 \def\slash#1{\setbox0\hbox{$ #1 $}
 \setbox0\hbox to \the\slashfraction\wd0{\hss \box0}/\box0 }
 \unlock
 \def\leftrightarrowfill{$\m@th\mathord-\mkern-6mu%
   \cleaders\hbox{$\mkern-2mu\mathord-\mkern-2mu$}\hfill
   \mkern-6mu\mathord\leftrightarrow$}
 \def\overlrarrow#1{\vbox{\ialign{##\crcr
       \leftrightarrowfill\crcr\noalign{\kern-\p@\nointerlineskip}
       $\hfil\displaystyle{#1}\hfil$\crcr}}}
 \lock
%

{\obeyspaces\global\let =\ }   
%
\def\papersize{       \hsize=35pc\vsize=50pc\hoffset=1cm\voffset=1.3cm
              \pagebottomfiller=0pc
              \skip\footins=\bigskipamount\normalspace}
\def\lettersize{\hsize=6.5in\vsize=8.5in\hoffset=0cm\voffset=1.6cm
              \pagebottomfiller=\letterbottomskip
              \skip\footins=\smallskipamount
              \multiply\skip\footins by 3
              \singlespace}
\papers
%
\catcode`\@=11 
\newif\ifletterstyle                
\letterstylefalse             
\def\letters{\lettersize\letterstyletrue
   \headline=\letterheadline \footline=\letterfootline
   \immediate\openout\labelswrite=\jobname.lab}
\def\iftpub{\afterassignment\iftp@b\toks@}
\def\iftp@b{\edef\n@xt{\Pubnum={UFIFT--HEP--\the\toks@}}\n@xt}
\let\pubnum=\iftpub
\expandafter\ifx\csname eightrm\endcsname\relax
    \let\eightrm=\ninerm \let\eightbf=\ninebf \fi
\catcode`\@=12 
%
   
 

\unlock
\def\eightpoint{\relax
    \textfont0=\eightrm          \scriptfont0=\eightrm
    \scriptscriptfont0=\fiverm
    \def\rm{\fam0 \eightrm \f@ntkey=0 }\relax
    \textfont1=\eighti           \scriptfont1=\eighti
    \scriptscriptfont1=\fivei
    \def\oldstyle{\fam1 \eighti \f@ntkey=1 }\relax
    \textfont2=\eightsy          \scriptfont2=\eightsy
    \scriptscriptfont2=\fivesy
    \textfont3=\tenex          \scriptfont3=\tenex
    \scriptscriptfont3=\tenex
    \def\it{\fam\itfam \eightit \f@ntkey=4 }\textfont\itfam=\eightit
    \def\sl{\fam\slfam \eightsl \f@ntkey=5 }\textfont\slfam=\eightsl
    \def\bf{\fam\bffam \eightbf \f@ntkey=6 }\textfont\bffam=\eightbf
        \scriptfont\bffam=\eightbf     \scriptscriptfont\bffam=\fivebf
    \def\tt{\fam\ttfam \eighttt \f@ntkey=7 }\textfont\ttfam=\eighttt
    \setbox\strutbox=\hbox{\vrule height 4pt depth 3pt width\z@}
    \samef@nt}
\lock
\def\boxit#1{\vbox{\hrule\hbox{\vrule\kern3pt
             \vbox{\kern3pt#1\kern3pt}\kern3pt\vrule}\hrule}}
 \newdimen\str

\def\fboxit#1#2{\vbox{\hrule height #1 \hbox{\vrule width #1
           \kern3pt \vbox{\kern3pt#2\kern3pt}\kern3pt \vrule width #1 }
           \hrule height #1 }}

\def\fillbox#1{\hbox to #1{\vbox to #1{\vfil}\hfil}}
\def\dotbox#1{\hbox to #1{\vbox to 10pt{\vfil}\hss $\cdots$ \hss}}
\def\ggenbox#1#2{\vbox to 10pt{\vss \hbox to #1{\hss #2  \hss} \vss}}


\catcode`\@=11 
\newtoks\foottokens
\let\labelfont=\Tenpoint      
\def\MakeFromBox{\gl@bal\setbox\FromLabelBox=\vbox{\labelfont
     \ialign{##\hfil\cr \the\sendername \the\FromAddress \crcr }}}
\def\smallsize{\relax
\def\eightpoint{\relax
\textfont0=\eightrm  \scriptfont0=\sixrm
\scriptscriptfont0=\fiverm
\def\rm{\fam0 \eightrm \f@ntkey=0}\relax
\textfont1=\eighti  \scriptfont1=\sixi
\scriptscriptfont1=\fivei
\def\oldstyle{\fam1 \eighti \f@ntkey=1}\relax
\textfont2=\eightsy  \scriptfont2=\sixsy
\scriptscriptfont2=\fivesy
\textfont3=\tenex  \scriptfont3=\tenex
\scriptscriptfont3=\tenex
    \def\it{\fam\itfam \eightit \f@ntkey=4 }\textfont\itfam=\eightit
\def\sl{\fam\slfam \eightsl \f@ntkey=5 }\textfont\slfam=\eightsl
\def\bf{\fam\bffam \eightbf \f@ntkey=6 }\textfont\bffam=\eightbf
\scriptfont\bffam=\sixbf   \scriptscriptfont\bffam=\sixbf
\def\tt{\fam\ttfam \eighttt \f@ntkey=7 }
\def\caps{\fam\cpfam \tencp \f@ntkey=8 }\textfont\cpfam=\tencp
\setbox\strutbox=\hbox{\vrule height 7.35pt depth 3.02pt width\z@}
\samef@nt}
\normalbaselineskip = 16.60pt plus 0.166pt minus 0.083pt
\normallineskip = 1.25pt plus 0.08pt minus 0.08pt
\normallineskiplimit = 1.25pt
\normaldisplayskip = 16.60pt plus 4.15pt minus 8.3pt
\normaldispshortskip = 4.98pt plus 3.32pt
\normalparskip = 4.98pt plus 1.67pt minus .83pt
\skipregister = 4.15pt plus 1.67pt minus 1.25pt
\def\Eightpoint{\eightpoint \relax
  \ifsingl@\subspaces@t2:5;\else\subspaces@t3:5;\fi
  \ifdoubl@ \multiply\baselineskip by 5
            \divide\baselineskip by 4\fi }
\parindent=16.67pt
\itemsize=25pt
\thinmuskip=2.5mu
\medmuskip=3.33mu plus 1.67mu minus 3.33mu
    \thickmuskip=4.17mu plus 4.17mu
\def\thinspace{\kern .13889em }
\def\negthinspace{\kern-.13889em }
\def\enspace{\kern.416667em }
\def\enskip{\hskip.416667em\relax}
\def\quad{\hskip.83333em\relax}
\def\qquad{\hskip1.66667em\relax}
\def\crr{\cropen{8.3333pt}}
\labelwidth=4.5in
\let\labelfont=\Eightpoint
\let\letterhead=\FLOHEAD      
\def\Vfootnote##1{\insert\footins\bgroup
   \interlinepenalty=\interfootnotelinepenalty \floatingpenalty=20000
   \singl@true\doubl@false\Eightpoint
   \splittopskip=\ht\strutbox \boxmaxdepth=\dp\strutbox
   \leftskip=\footindent \rightskip=\z@skip
   \parindent=0.5\footindent \parfillskip=0pt plus 1fil
   \spaceskip=\z@skip \xspaceskip=\z@skip \footnotespecial
   \Textindent{##1}\footstrut\futurelet\next\fo@t}%
\def\attach##1{\step@ver{\strut^{\mkern 1.6667mu ##1} } }
\def\inserttable ##1##2##3%
    {%
    \tbldef {##1}{##3}\goodbreak%
        \midinsert
      \smallskip
      \hbox{\singlespace \hskip 0.5cm
              \vtop{\parshape=2 0cm 10.8cm 1.3cm 9.5cm
                    \noindent{\bf\Table{##1}}.\enspace ##3}
              \hfil}
      ##2
      \smallskip
    \endinsert
    }
\def\sure{y}
\def\insertfigure ##1##2##3%
    {%
    \figdef {##1}{##3}\goodbreak%
    \midinsert
      \smallskip
      ##2
      \hbox{\singlespace\hskip 0.5cm
              \vtop{\parshape=2 0cm 10.8cm
                      1.6cm 9.2cm \noindent{\bf\Figure{##1}}.
                      \enspace ##3}
              \hfil}
        \smallskip
    \endinsert
    }%
\def\references{\par\penalty-300\vskip\chapterskip
        \spacecheck\chapterminspace
      \line{\twelverm\hfil REFERENCES\hfil}
      \nobreak\vskip\headskip\penalty 30000
      \reflist{}}
\def\figures{\par\penalty-300\vskip\chapterskip
        \spacecheck\chapterminspace
      \line{\twelverm\hfil FIGURE CAPTIONS\hfil}
      \nobreak\vskip\headskip\penalty 30000
      \figlist{}}
\def\tables{\par\penalty-300\vskip\chapterskip
        \spacecheck\chapterminspace
      \line{\twelverm\hfil TABLE CAPTIONS\hfil}
      \nobreak\vskip\headskip\penalty 30000
      \tbllist{}}
\def\PH@SR@V{\doubl@true\baselineskip=20.08pt plus .1667pt minus .0833pt
             \parskip = 2.5pt plus 1.6667pt minus .8333pt }
    \def\author##1{\vskip\frontpageskip\titlestyle{\tencp ##1}\nobreak}
\def\address##1{\par\kern 4.16667pt\titlestyle{\tenpoint\it ##1}}
\def\andaddress{\par\kern 4.16667pt \centerline{\sl and} \address}
\def\UFL{\address{Department of Physics\break
      University of Florida, Gainesville, FL 32611}}
\def\abstract{\vskip\frontpageskip\centerline{\twelverm ABSTRACT}
              \vskip\headskip }
\def\submit##1{\par\nobreak\vfil\nobreak\medskip
   \centerline{Submitted to \sl ##1}}
\def\doeack{\foot{Work supported in part by the Department of Energy
              under grant  DE--FG05--86ER--40272.}}
\def\nsfack{\foot{Work supported by National Science Foundation
              Grant  PHY 84--16030A01.}}
\def\cases##1{\left\{\,\vcenter{\Tenpoint\m@th
    \ialign{$####\hfil$&\quad####\hfil\crcr##1\crcr}}\right.}
\def\matrix##1{\,\vcenter{\Tenpoint\m@th
    \ialign{\hfil$####$\hfil&&\quad\hfil$####$\hfil\crcr
      \mathstrut\crcr\noalign{\kern-\baselineskip}
     ##1\crcr\mathstrut\crcr\noalign{\kern-\baselineskip}}}\,}
\Tenpoint
    }
\newdimen\fullhsize
\newbox\leftcolumn
\def\twoinone{
\smallsize
\def\papersize{
              \voffset=-.23truein
              \vsize=7truein
              \baselineskip=16pt plus 2pt minus 1pt
              \fullhsize=10truein\hsize=4.75truein
                \hoffset=-.54truein
              \skip\footins=\bigskipamount}
\def\lettersize{\voffset=.31truein
              \vsize=6.38truein
              \baselineskip=16pt plus 2pt minus 1pt
              \fullhsize=10truein\hsize=4.75truein
                \hoffset=-.48truein
              \skip\footins=\smallskipamount
                \multiply\skip\footins by3}
\papers               
    \let\lr=L
\output={\if L\lr
              \global\setbox\leftcolumn=\columnbox \advancepageno
              \global\let\lr=R
       \else  \getitout \advancepageno
              \global\let\lr=L\fi
       \ifnum\outputpenalty>-20000 \else\dosupereject\fi}
}             
      \def\columnbox{\leftline{
              \vbox{\ifletterstyle\makeheadline\fi
                      \pagebody\makefootline}}}
      \def\fullline{\hbox to\fullhsize}
      \def\getitout{\shipout\vbox{\fullline{\box\leftcolumn
              \hfil {\leftline{
              \vbox{\makeheadline
              \pagebody\makefootline}}} }}}
%
\catcode`\@=12 
    %
%
%
\newcount      \ObjClass
\chardef\ClassNum     = 0
\chardef\ClassMisc    = 1
\chardef\ClassEqn     = 2
\chardef\ClassRef     = 3
\chardef\ClassFig     = 4
\chardef\ClassTbl     = 5
\chardef\ClassThm     = 6
\chardef\ClassStyle     = 7
    \chardef\ClassDef       = 8
\edef\NumObj  {\ObjClass = \ClassNum   \relax}
\edef\MiscObj {\ObjClass = \ClassMisc  \relax}
\edef\EqnObj  {\ObjClass = \ClassEqn   \relax}
\edef\RefObj  {\ObjClass = \ClassRef   \relax}
\edef\FigObj  {\ObjClass = \ClassFig   \relax}
\edef\TblObj  {\ObjClass = \ClassTbl   \relax}

\edef\StyleObj  {\ObjClass = \ClassStyle \relax}
\edef\DefObj    {\ObjClass = \ClassDef   \relax}
%
%
\def\gobble    #1{}%
\def\trimspace   #1 \end{#1}%
\def\ifundefined #1{\expandafter \ifx \csname#1\endcsname \relax}%
\def\trimprefix  #1_#2\end{\expandafter \string \csname #2\endcsname}%
\def\skipspace #1#2#3\end%
    {%
    \def \temp {#2}%
    \ifx \temp\space \skipspace #1#3\end
    \else \gdef #1{#2#3}\fi
        }%
\def\stylename#1{\expandafter\expandafter\expandafter
    \gobble\expandafter\string\the#1}
\ifundefined {protect} \let\protect=\relax \fi
\catcode`\@=11
\let\rel@x=\relax
\def\relaxtest{\rel@x}
\catcode`\@=12
\def\checkchapterlabel%
    {%
        {\protect\if\chapterlabel\relaxtest
      \global\let\chapterlabel=\relax\fi}
    }%
\begingroup
\catcode`\<=1 \catcode`\{=12
\catcode`\>=2 \catcode`\}=12
\xdef\LBrace<{>%
\xdef\RBrace<}>%
\endgroup
%
%
\newcount\equanumber \equanumber=0
    \newcount\eqnumber   \eqnumber=0
\newif\ifleftnumbers \leftnumbersfalse

\def\(#1)%
     {%
        \ifnum \equanumber<0 \eqnumber=-\equanumber
          \advance\eqnumber by -1 \else
            \eqnumber = \equanumber\fi
        \ifmmode\ifinner(\eqnum{#1})\else
        \ifleftnumbers\leqno(\eqnum{#1})\ifdraft{\rm[#1]}\fi
            \else\eqno(\eqnum{#1})\ifdraft{\rm[#1]}\fi\fi\fi
      \else(\eqnum{#1})\fi\ifnum%
          \equanumber<0 \global\equanumber=-\eqnumber\global\advance
            \equanumber by -1\else\global\equanumber=\eqnumber\fi
     }%
\def\mideq(#1)%
     {%
      \ifleftnumbers \leqinsert{$\(#1)$} \else
      \eqinsert{$\(#1)$} \fi
     }%
\def\eqnum #1%
    {%
    \LookUp Eq_#1 \using\eqnumber\neweqnum
    {\rm\label}%
    }%
\def\neweqnum #1#2%
    {%
    \checkchapterlabel
    {\protect\xdef\eqnoprefix{\ifundefined{chapterlabel}
      \else\chapterlabel.\fi}}
    \ifmmode \xdef #1{\eqnoprefix #1}
        \else\message{Undefined equation \string#1 in non-math mode.}%
           \xdef #1{\relax}
           \global\advance \eqnumber by -1
        \fi
    \EqnObj \SaveObject{#1}{#2}
    }%
\everydisplay = {\expandafter \let\csname Eq_\endcsname=\relax
               \expandafter \let\csname Eq_?\endcsname=\relax}%
%
    %
\newcount\tablecount \tablecount=0
\def\Table  #1{Table~\tblnum {#1}}%
\def\tblnum #1{\TblObj \LookUp Tbl_#1 \using\tablecount
      \SaveObject \label\ifdraft [#1]\fi}%
\def\tbldef #1{\TblObj \SaveContents {Tbl_#1}}%
\def\tbllist  {\TblObj \ListObjects}%
%
\def\inserttable #1#2#3%
    {%
    \tbldef {#1}{#3}\goodbreak%
    \midinsert
      \smallskip
      \hbox{\singlespace
            \vtop{\titlestyle{{\Tenpoint{\caps\Table{#1}}\break #3}}}
           }%
      #2
      \smallskip%
    \endinsert
    }%
    \def\topinserttable #1#2#3%
    {%
    \tbldef {#1}{#3}\goodbreak%
    \topinsert
      \smallskip
      \hbox{\singlespace
            \vtop{\titlestyle{{\Tenpoint{\caps\Table{#1}}\break #3}}}
           }%
      #2
      \smallskip%
    \endinsert
    }%
%
%
\newcount\figurecount \figurecount=0
\def\Figure #1{Figure~\fignum {#1}}%
\def\Fig    #1{Fig.~\fignum {#1}}%
\def\fignum #1{\FigObj \LookUp Fig_#1 \using\figurecount
     \SaveObject \label\ifdraft [#1]\fi}%
    \def\figdef #1{\FigObj \SaveContents {Fig_#1}}%
\def\figlist  {\FigObj \ListObjects}%
%
\def\sure{y}
\def\insertfigure #1#2#3#4%
    {%
    \figdef {#1}{#3}%
    \midinsert
      \bigskip
      \ifx\havefigures\sure
      #2
      \else
         {#4}\fi
      \hbox{  \singlespace
              \hskip 0.4in
              \vtop{\parshape=2 0pt 362pt 32pt 330pt
                    \noindent{\Tenpoint{\caps\Fig{#1}}.\enspace #3}}
              \hfil}
      \smallskip%
    \endinsert
    }%
\def\topinsertfigure #1#2#3%
    {%
        \figdef {#1}{#3}%
    \topinsert
      \bigskip
      #2
      \hbox{  \singlespace
              \hskip 0.4in
              \vtop{\parshape=2 0pt 362pt 32pt 330pt
                    \noindent{\Tenpoint{\caps\Fig{#1}}.\enspace #3}}
              \hfil}
      \smallskip%
    \endinsert
    }%
%
%
%
    \newcount\theoremcount \theoremcount=0
%
%
%
%
%
%
%
%
%
%
%
%
%
%
\newcount\referencecount \referencecount=0
\newcount\refsequence \refsequence=0
\newcount\lastrefno   \lastrefno=-1
%
\def\NPrefs{\let\refmark=\NPrefmark \let\refitem=\NPrefitem}

%
\def\refsymbol#1{\refrange#1-\end}%
\def\[#1]#2%
      {%
      \if.#2\rlap.\refmark{\refsymbol{#1}}\let\refendtok=\relax%
      \else\if,#2\rlap,\refmark{\refsymbol{#1}}\let\refendtok=\relax%
      \else\refmark{\refsymbol{#1}}\let\refendtok=#2\fi\fi%
      \discretionary{}{}{}\refendtok}%
\def\refrange #1-#2\end%
    {%
    \refnums #1,\end
    \def \temp {#2}%
    \ifx \temp\empty \else -\expandafter\refrange \temp\end \fi
    }%
\def\refnums #1,#2\end%
    {%
    \def \temp {#1}%
    \ifx \temp\empty \else \skipspace \temp#1\end\fi
    \ifx \temp\empty
      \ifcase \refsequence
          \or\or ,\number\lastrefno
            \else  -\number\lastrefno
      \fi
      \global\lastrefno = -1
      \global\refsequence = 0
    \else
      \RefObj \edef\temp {Ref_\temp\space}%
      \expandafter \LookUp \temp \using\referencecount\SaveObject
      \global\advance \lastrefno by 1
      \edef \temp {\number\lastrefno}%
      \ifx \label\temp
          \global\advance\refsequence by 1
      \else
          \global\advance\lastrefno by -1
          \ifcase \refsequence
              \or ,%
              \or ,\number\lastrefno,%
          \else   -\number\lastrefno,%
          \fi
          \label
          \global\refsequence = 1
          \ifx\suffix\empty
              \global\lastrefno = \label
          \else
                \global\lastrefno = -1
          \fi
      \fi
      \refnums #2,\end
    \fi
    }%
%
%
%
%
%
\def\reflist  {\RefObj \ListObjects}%
\def\Refer #1{Ref.[\refsymbol{#1}]}%
%
%
\newif\ifSaveFile
\newif\ifnotskip
\newwrite\SaveFile
    \let\IfSelect=\iftrue
\edef\savefilename {\jobname.aux}%
\def\Def#1#2%
    {%
    \expandafter\gdef\noexpand#1{#2}%
    \DefObj \SaveObject {#2}{\expandafter\gobble\string#1}%
}%
\def\savestate%
    {%
    \ifundefined {chapternumber} \else
      \NumObj \SaveObject {\number\chapternumber}{chapternumber} \fi
        \ifundefined {appendixnumber} \else
      \NumObj \SaveObject {\number\appendixnumber}{appendixnumber} \fi
      \ifundefined {sectionnumber} \else
      \NumObj \SaveObject {\number\sectionnumber}{sectionnumber} \fi
    \ifundefined {pagenumber} \else
      \advance\pagenumber by 1
      \NumObj \SaveObject {\number\pagenumber}{pagenumber}%
      \advance\pagenumber by -1 \fi
    \NumObj \SaveObject {\number\equanumber}{equanumber}%
    \NumObj \SaveObject {\number\tablecount}{tablecount}%
        \NumObj \SaveObject {\number\figurecount}{figurecount}%
    \NumObj \SaveObject {\number\theoremcount}{theoremcount}%
    \NumObj \SaveObject {\number\referencecount}{referencecount}%
    \checkchapterlabel
    \ifundefined {chapterlabel} \else
      {\protect\xdef\chaplabel{\chapterlabel}}
      \MiscObj \SaveObject \chaplabel {chapterlabel} \fi
    \ifundefined {chapterstyle} \else
      \StyleObj \SaveObject
           {\stylename{\chapterstyle}}{chapterstyle} \fi
    \ifundefined {appendixstyle} \else
      \StyleObj \SaveObject
           {\stylename{\appendixstyle}}{appendixstyle}\fi
}%
\def\Contents #1{\ObjClass=-#1 \SaveContents}%
\def\Define #1#2#3%
    {%
    \ifnum #1=\ClassNum
      \global \csname#2\endcsname = #3 %
    \else \ifnum #1=\ClassStyle
      \global \csname#2\endcsname\expandafter=
        \expandafter{\csname#3\endcsname} %
    \else \ifnum #1=\ClassDef
        \expandafter\gdef\csname#2\endcsname{#3} %
    \else
      \expandafter\xdef \csname#2\endcsname {#3} \fi\fi\fi %
    \ObjClass=#1 \SaveObject {#3}{#2}%
    }%
\def\SaveObject #1#2%
    {%
    \ifSaveFile \else \OpenSaveFile \fi
    \immediate\write\SaveFile
      {%
      \noexpand\IfSelect\noexpand\Define
      {\the\ObjClass}{#2}{#1}\noexpand\fi
      }%
    }%
\def\SaveContents #1%
    {%
    \ifSaveFile \else \OpenSaveFile \fi
    \BreakLine
    \SaveLine {#1}%
    }%
\begingroup
        \catcode`\^^M=\active %
\gdef\BreakLine %
    {%
    \begingroup %
    \catcode`\^^M=\active %
    \newlinechar=`\^^M %
    }%
\gdef\SaveLine #1#2%
    {%
    \toks255={#2}%
    \immediate\write\SaveFile %
      {%
      \noexpand\IfSelect\noexpand\Contents
      {-\the\ObjClass}{#1}\LBrace\the\toks255\RBrace\noexpand\fi%
      }%
    \endgroup %
    }%
\endgroup
\def\ListObjects #1%
    {%
    \ifSaveFile \CloseSaveFile \fi
    \let \IfSelect=\GetContents \ReadFileList #1,\savefilename,\end
       \let \IfSelect=\IfDoObject  \input \savefilename
    \let \IfSelect=\iftrue
    }%
\def\ReadFileList #1,#2\end%
    {%
    \def \temp {#1}%
    \ifx \temp\empty \else \skipspace \temp#1\end \fi
    \ifx \temp\empty \else \input #1 \fi
    \def \temp {#2}%
    \ifx \temp\empty \else \ReadFileList #2\end \fi
    }%
\def\GetContents #1#2#3%
    {%
    \notskipfalse
    \ifnum \ObjClass=-#2
      \expandafter\ifx \csname #3\endcsname
        \relax \else \notskiptrue \fi
    \fi
    \ifnotskip \expandafter \DefContents \csname #3_\endcsname
    }%
\def\DefContents #1#2{\toks255={#2} \xdef #1{\the\toks255}}%
    \def\IfDoObject #1#2%
    {%
    \notskipfalse \ifnum \ObjClass=#2
        \notskiptrue\fi \ifnotskip \DoObject
    }%
\def\DoObject #1#2%
    {%
    \ifnum \ObjClass = \ClassTbl      \par\noindent Table~#2.
    \else \ifnum \ObjClass = \ClassFig        \par\noindent Figure~#2.
    \else \ifnum \ObjClass = \ClassRef  \refitem{#2}
    \else \item {#2.}
    \fi\fi\fi
    \ifdraft\edef\temp
       {\trimprefix #1\end}[\expandafter\gobble \temp]~\fi
    \expandafter\ifx \csname #1_\endcsname \relax
      \ifdraft\relax\else\edef\temp {\trimprefix #1\end}%
      [\expandafter\gobble \temp]\fi%
    \else
      \csname #1_\endcsname
    \fi
    }%
\def\OpenSaveFile   {\immediate\openout\SaveFile=\savefilename
                   \global\SaveFiletrue}%
\def\CloseSaveFile  {\immediate\closeout
                 \SaveFile \global\SaveFilefalse}%
%
%
\def\LookUp #1 #2\using#3#4%
    {%
    \expandafter \ifx\csname#1\endcsname \relax
        \global\advance #3 by 1
        \expandafter \xdef \csname#1\endcsname {\number #3}%
        \let \newlabelfcn=#4%
          \ifx \newlabelfcn\relax \else
           \expandafter \newlabelfcn \csname#1\endcsname {#1}%
         \fi
    \fi
    \xdef \label  {\csname#1\endcsname}%
    \gdef \suffix {#2}%
    \ifx \suffix\empty \else
        \xdef \suffix {\expandafter\trimspace \suffix\end}%
        \xdef \label  {\label\suffix}%
    \fi
    }%
   %
%
%
\newcount\appendixnumber        \appendixnumber=0
\newtoks\appendixstyle                \appendixstyle={\Alphabetic}
\newif\ifappendixlabel                \appendixlabelfalse
\def\APPEND#1{\par\penalty-300\vskip\chapterskip
      \spacecheck\chapterminspace
      \global\chapternumber=\number\appendixnumber
      \global\advance\appendixnumber by 1
      \chapterstyle\expandafter=\expandafter{\the\appendixstyle}
\chapterreset
     \titlestyle{Appendix\ifappendixlabel~\chapterlabel\fi.~ {#1}}
      \nobreak\vskip\headskip\penalty 30000}
%

%
%
%
\def\references#1{\par\penalty-300\vskip\chapterskip\spacecheck
       \chapterminspace\line{\fourteenrm\hfil References\hfil}
       \nobreak\vskip\headskip\penalty 30000\reflist{#1}}
\def\figures#1{\par\penalty-300\vskip\chapterskip\spacecheck
     \chapterminspace\line{\fourteenrm\hfil Figure Captions\hfil}
     \nobreak\vskip\headskip\penalty 30000\figlist{#1}}
\def\tables#1{\par\penalty-300\vskip\chapterskip\spacecheck
      \chapterminspace\line{\fourteenrm\hfil Table Captions\hfil}
       \nobreak\vskip\headskip\penalty 30000\tbllist{#1}}
\newif\ifdraft\draftfalse
\newcount\yearltd\yearltd=\year\advance\yearltd by -1900
\def\draft{\drafttrue
    \def\draftdate{preliminary draft:
        \number\month/\number\day/\number\yearltd\ \ \hourmin}%
        \paperheadline={\hfil\draftdate} \headline=\paperheadline
        {\count255=\time\divide\count255 by 60
                   \xdef\hourmin{\number\count255}
            \multiply\count255 by-60\advance\count255 by\time
    \xdef\hourmin{\hourmin:\ifnum\count255<10 0\fi\the\count255} }
      \message{draft mode}  }
%
\def\slacpub{ \Pubnum={$\caps SLAC - PUB - \the\pubnum $}}
%
%

\def\frac#1/#2{\leavevmode\kern.1em\raise.5ex
              \hbox{\the\scriptfont0
              #1}\kern-.1em/\kern-.15em
              \lower.25ex\hbox{\the\scriptfont0 #2}}
%
%

\def\frac#1#2{{#1 \over #2}}

\def\half{{\frac 12}}

\def\fourth{{\frac 14}}
\def\balpha{\hbox{\twelvembf\char\number 11}}
\def\bgamma{\hbox{\twelvembf\char\number 13}}

%

%
\IfSelect \Contents
{-3}{Ref_thornmosc}{C. B. Thorn,
``Reformulating String Theory with the 1/N Expansion,''
in {\it Sakharov Memorial Lectures in Physics},
Ed. L. V. Keldysh and V. Ya. Fainberg, Nova Science Publishers Inc.,
Commack, New York, 1992. hep-th/9405069}\fi
\IfSelect \Contents
{-3}{Ref_klebanovs}{I. Klebanov and L. Susskind,
{\sl Nucl. Phys.} {\bf B309} (1988) 175.}\fi
\IfSelect \Contents
{-3}{Ref_thooftlargen}{G. 't Hooft,
{\sl Nucl. Phys.} {\bf B72} (1974) 461.}\fi
\IfSelect \Contents
{-3}{Ref_thooftbh}{G. 't Hooft,
{\sl Nucl. Phys.} {\bf B342} (1990) 471;
``On the Quantization of Space and Time,'' {\it Proc. of the
4th Seminar on Quantum Gravity}, 25\dash29 May 1987, Moscow, USSR,
ed. M. A. Markov \etal, (World Scientific Press, 1988);
``Dimensional Reduction in Quantum Gravity,'' Utrecht preprint,
THU-93/26, GR-QC/9310026.}\fi
\IfSelect \Contents
{-3}{Ref_nielsenfishnet}{H. B. Nielsen and P. Olesen,
{\sl Phys. Lett.} {\bf 32B} (1970) 203;
B. Sakita and M. A. Virasoro,
{\sl Phys. Rev. Lett.} {\bf 24} (1970) 1146.}\fi
\IfSelect \Contents
{-3}{Ref_gilest}{R. Giles and C. B. Thorn,
{\sl Phys. Rev.} {\bf D16} (1977) 366.}\fi
\IfSelect \Contents
{-3}{Ref_thornfishnet}{C. B. Thorn,
{\sl Phys. Rev.} {\bf D17} (1978) 1073.}\fi
\IfSelect \Contents
{-3}{Ref_bardakcis}{K. Bardakci and S. Samuel,
{\sl Phys. Rev.} {\bf D16} (1977) 2500.}\fi
\IfSelect \Contents
{-3}{Ref_gilesmt}{R. Giles, L. McLerran, C. B. Thorn,
{\sl Phys. Rev.} {\bf D17} (1978) 2058.}\fi
\IfSelect \Contents
{-3}{Ref_thornfock}{C. B. Thorn,
{\sl Phys. Rev.} {\bf D20} (1979) 1435.}\fi
\IfSelect \Contents
{-3}{Ref_thornrpa}{C. B. Thorn,
{\sl Phys. Rev.} {\bf D51} (1995) 647.}\fi
\IfSelect \Contents
{-3}{Ref_thornweeparton}{C. B. Thorn,
{\sl Phys. Rev.} {\bf D19} (1979) 639.}\fi
\IfSelect \Contents
{-3}{Ref_puzalowski}{R. Puzalowski,
{\sl Acta Physica Austriaca} {\bf 50} (1978) 45.}\fi
\IfSelect \Contents
{-3}{Ref_feynmanparton}{R. P. Feynman,
Third Topical Conference in High Energy Collisions
of Had\-rons, Stony Brook, N.Y.(1969);
J. D. Bjorken and E. Paschos,
{\sl Phys. Rev.} {\bf 185} (1969) 1975;
J. Kogut and L. Susskind, {\sl Physics Reports} {\bf 8} (1973) 75.}\fi
\IfSelect \Contents
{-3}{Ref_susskindbh}{L. Susskind,
``Some Speculations About Black Hole Entropy in String Theory,''
Rutgers Univ. preprint RU-93-44, hep-th/9309145;
L. Susskind and J. Uglum, ``Black Hole Entropy in Canonical
Quantum Gravity and Superstring Theory,'' Stanford Univ. preprint,
hep-th/9401070;
L. Susskind,
{\sl Phys. Rev.} {\bf D49} (1994) 6606.}\fi
\IfSelect \Contents
{-3}{Ref_bergmantgal}{O. Bergman and C. B. Thorn,
``Super-Galilei Invariant Field Theories in 2 $+$ 1 Dimensions,''}\fi
\IfSelect \Contents
{-3}{Ref_greenssustring}{M. B. Green and J. H. Schwarz,
{\sl Phys. Lett.} {\bf B109} (1982) 444;
{\sl Nucl. Phys.} {\bf B181} (1981) 502;
{\sl Nucl. Phys.} {\bf B198} (1982) 252.}\fi
\IfSelect \Contents
{-3}{Ref_greensb}{M. B. Green, J. H. Schwarz, and L. Brink,
{\sl Nucl. Phys.} {\bf B219} (1983) 437.}\fi
\IfSelect \Contents
{-3}{Ref_greensbook}{M. B. Green, J. H. Schwarz, and E. Witten,
{\it Superstring Theory}, Volumes 1 and 2,
Cambridge University Press (1987).}\fi
\IfSelect \Contents
{-3}{Ref_greenseiberg}{M. B. Green and N. Seiberg,
{\sl Nucl. Phys.} {\bf B299} (1988) 108.}\fi
\IfSelect \Contents
{-3}{Ref_greensiteklink}{J. Greensite and F. R. Klinkhamer,
{\sl Nucl. Phys.} {\bf B281} (1987) 269;
{\sl Nucl. Phys.} {\bf B291} (1987) 557;
{\sl Nucl. Phys.} {\bf B304} (1988) 108.}\fi
\def\sgone{{\cal S}_1{\cal G}}
\def\sgtwo{{\cal S}_2{\cal G}}

\def\bx{{\bf x}}
\def\by{{\bf y}}
\def\bz{{\bf z}}

\endpage
\pubnum={95--8\cr
hep-th/9506125\cr}
\date{}
\pubtype={}
\titlepage
\title{String Bit Models for Superstring\foot{Supported in part by
the Department of Energy under grant DE-FG05-86ER-40272,
and by the Institute for Fundamental Theory.}}
\author{Oren Bergman\foot{E-mail  address: oren@phys.ufl.edu \hfill}
and Charles B. Thorn\foot{E-mail  address: thorn@phys.ufl.edu \hfill}
}
\address{Institute for Fundamental Theory\break
Department of Physics, University of Florida, Gainesville,
FL, 32611, USA }
\abstract
\noindent We extend the model of string as a polymer of string bits
to the case of
superstring. We mainly concentrate
on
type II-B  superstring, with some discussion of
the obstacles presented by not II-B superstring, together
with possible strategies for surmounting them.
As with previous work on
bosonic string we work within the light-cone gauge.
The bit model possesses a good deal less symmetry than the
continuous string theory. For one thing, the bit model is
formulated as a Galilei invariant theory in $(D-2)+1$ dimensional
space-time. This means that
Poincar\'e invariance
is reduced to the
Galilei subgroup in $D-2$ space
dimensions. Naturally the supersymmetry present in the bit
model is likewise dramatically reduced. Continuous string can
arise in the bit models with the formation of infinitely long
polymers of string bits. Under the right circumstances
(at the critical dimension) these
polymers can behave as string moving in
$D$ dimensional space-time
enjoying the full $N=2$ Poincar\'e supersymmetric dynamics of
type II-B superstring.
\endpage
\pagenumbers
\chapter{Introduction}
The idea that relativistic string is a composite of point
like entities\[gilest,klebanovs,thornmosc]
called ``string bits'' is an appealing alternative
to the cumbersome formal apparatus of string field theory.
The origins of the idea can be traced to the earliest days
of dual models\[nielsenfishnet] with the attempt, motivated
in part by the old parton model of hadrons,\[feynmanparton]
to understand dual resonance amplitudes
as planar ``fishnet'' Feynman diagrams.
After 't Hooft showed that planar diagrams are naturally
singled out by the $1/N_c$ expansion,\[thooftlargen] the idea was again
vigorously explored as a possible link between nonabelian
gauge theory
and string theory\[thornfishnet,bardakcis,thornweeparton].
\hskip-3pt
The attempted linkage failed because, unlike the partons of
hadrons (quarks and gluons), the ``partons'' of string
never carry a finite fraction of the string's momentum:
string bits are always ``wee'' partons.
{}From the modern point of view, strings are not hadrons and
we advocate that the inevitable weeness of string bits should
actually be embraced as a uniquely stringy hallmark\[thornmosc].

Our main goal in developing string bit models
is to devise
a truly nonperturbative formulation of string
theory. In the earlier work of one of us
this idea has been pursued only in light-cone
gauge and systematically developed
only for
bosonic string\[thornrpa].
Bosonic string
(in 26 space-time dimensions)
is generally believed to be absolutely unstable, and it is therefore
an unfortunate test case for a nonperturbative reformulation.
This has not hindered the formal implementation of
string bit ideas for this case, since that has so far
been limited to a perturbative
context. However there
seems little point in attempting nonperturbative studies of
bosonic string bit models, other than to confirm that
they don't make sense as string theories.
We can be much more optimistic in the case of
superstring theory
which is generally hoped to be a consistent stable
theory. Indeed, if a superbit model
for superstring can be shown to be a good theory at the
nonperturbative level, there is the exciting possibility that
many of the conundrums of quantum gravity, such as the
consistency of quantum mechanics
in the presence of black holes
may be resolved\[thooftbh,susskindbh].

In this paper we present a bit model for
superstring, restricting attention for the most part to the
type II-B case, which presents the fewest obstacles to
a complete treatment.
By no means do we claim that our bit model is
unique. Universality suggests that the model can be generalized
in various ways, and still yield a satisfactory continuum limit. In fact
to get the correct string interactions the model {\bf has} to be
extended. Producing one or another satisfactory model is useful
for studying superstring theory, but we eventually want to restrict
the models by some underlying symmetry principles, not by whether
they possess a satisfactory continuum limit. Our bit model suggests
what some of these principles may be, but it certainly does not
give them all.

A dramatic feature of string theory viewed in light-cone gauge
is the fact that the longitudinal coordinate $x^-=(t-z)/\sqrt2$
is virtually eliminated from the theory. Except for its
zero mode, conjugate to $P^+$, it is solely a function of the
transverse coordinates. The string bit idea effectively
eliminates even this zero-mode longitudinal degree of freedom,
by identifying $P^+$ with the number of  string bits: each bit
is free to move around only in the transverse space. The
full space-time symmetry group of the string bit dynamics is
the
Galilei group in $(D-2)+1$ dimensional space-time
with  space coordinates $x^k$, $k=1,\cdots,D-2$ and time identified
with $x^+$. Each bit has a fixed Newtonian mass $m$. If $M$ bits
can form into long polymers, then $mM$ can be identified in the
limit $M\rightarrow\infty$ as the string's total $P^+$. All of
this has already been discussed in the simplified context of
bosonic string.\[gilest,thornmosc,thornweeparton,thornrpa,thornfock]
To extend the work to
superstring, we must decide
how the world-sheet spinors are to be fit into the string
bit picture. We shall find that they can emerge in the
continuous string limit if each bit is in a  256
component supermultiplet of $\sgone$,
the  minimal Super-Galilei
group\[puzalowski,bergmantgal]\ for 8 dimensional space.

The paper is organized as follows. In Section 2 we review the
Super-Poincar\'e algebra in light-cone coordinates and display
its Super-Galilei subalgebra. Then in section 3 we devise
a suitable discretization of
superstring in the light-cone
Green-Schwarz formulation.
This discretization motivates our
proposal for a fully second-quantized superstring bit model. In section
4 we present such models, first in $2+1$ dimensions as a warmup, then in
$8+1$ dimensions for
type II-B superstring.
Section 5 contains our
concluding remarks, which include a brief discussion of the
open issues
we leave for resolution in future work.
\chapter{Super-Poincar\'e Algebra in Light-Cone Coordinates}
We begin by reviewing the $D$-dimensional Super-Poincar\'e algebra and
expressing it in light-cone variables.
For simplicity we shall only consider even $D$.
The Super-Poincar\'e generators include
a vector $P^\mu$, a rank two antisymmetric tensor $M^{\mu\nu}$, and a
Grassmann odd spinor
$Q_{\cal A}$. Greek indices take values from $0$ to $D-1$, and
capital script indices
take values from $1$ to $2^{D/2}$, which is the dimension of the spinor
representation of the Poincar\'e group $ISO(D-1,1)$.
The algebra satisfied by
the generators is
given by
$$\eqalign{
[P^\mu,P^\nu] &= [Q_{\cal A},P^\mu] = 0 \cr
[M^{\mu\nu},P^\rho] &= i\big(\eta^{\mu\rho}P^\nu
- \eta^{\nu\rho}P^\mu\big) \cr
[M^{\mu\nu},M^{\rho\sigma}] &= i\big(\eta^{\mu\rho}M^{\nu\sigma}
          + \eta^{\mu\sigma}M^{\rho\nu} - \eta^{\nu\rho}M^{\mu\sigma}
          - \eta^{\nu\sigma}M^{\rho\mu}\big) \cr
[M^{\mu\nu},Q_{\cal A}] &=
- {1\over 2}\big(\Sigma^{\mu\nu}\cdot Q\big)_{\cal A} \cr
\{Q^{\phantom{\dagger}}_{\cal A},Q^{\dagger}_{\cal B}\}
&= - {1\over\sqrt2}
                           (\Gamma\cdot P\Gamma_0)_{\cal AB}\; ,\cr}
\(superpoincare)$$
where $\eta^{\mu\nu}={\rm diag}\{-1,1,\ldots,1\}$,
$\Gamma^\mu$ are the Dirac gamma matrices in $D$ dimensions,
and $\Sigma^{\mu\nu}={i\over 2}
[\Gamma^\mu,\Gamma^\nu]$. Note that the r.h.s. of the last
equation involves
$$-\Gamma\cdot P\Gamma^0=P^0+{P^k}
{\alpha^k},$$
where ${\alpha^k}\equiv\Gamma^0{\Gamma^k}$, $k=1,\cdots, D-1$,
are the original hermitian alpha matrices introduced by Dirac.

Light-cone coordinates are defined by singling out one of the spatial
directions,
say $x^{D-1}$, and letting
$$x^\pm \equiv {1\over\sqrt{2}}\big(x^0 \pm x^{D-1}\big) \; .
\(lightcone)$$
The role of time is played by $x^+$, so its
conjugate momentum $P^-$ plays the
role of the light-cone Hamiltonian. The longitudinal
coordinate is $x^-$, and the
transverse coordinates are $x^i$, with $i=1,\ldots ,D-2$.
In these coordinates a $(D-2)+1$ dimensional
Super-Galilei algebra
emerges as a sub-algebra of the full $D$ dimensional Super-Poincar\'e
algebra
in the transverse + time directions.
Transverse spatial translations are generated by $P^i$,
time translation is
generated by
$P^-$, transverse spatial rotations by $M^{ij}$,
and transverse Galilei boosts
by $M^{+i}$.
Accordingly, we make the replacements:
$$\eqalign{
 P^- &\rightarrow H\cr
 M^{ij} &\rightarrow J^{ij}\cr
 M^{+i} &\rightarrow K^i\; .\cr} \(replace)$$
The part of Super-Galilei sub-algebra involving even
generators is then given by
$$\eqalign{
[P^i,P^j] &= [P^i,H] = [J^{ij},H] = [K^i,K^j]=0\cr
[J^{ij},P^k] &= i\big(\delta^{ik}P^j - \delta^{jk}P^i\big) \cr
[K^i,P^j] &= -i\delta^{ij}P^+ \cr
[K^i,H] &= -iP^i \cr
[J^{ij},J^{kl}] &= i\big(\delta^{ik}J^{jl} + \delta^{il}J^{kj} -
\delta^{jk}J^{il}
                   -\delta^{jl}J^{ki} \big)\cr
[J^{ij},K^k] &= i\big(\delta^{ik}K^j - \delta^{jk}K^i\big) \; . \cr
}
\(supergalileo)$$
Note that in the above algebra
$P^+$ plays the role of the
Newtonian
mass. This role will be exploited in
constructing the string bit model for
discretized light-cone superstring,
in which $P^+$ is the length of
a piece of string, and is equal to the total
Newtonian mass of all the string bits.
The rest of the charges completing
the Super-Poincar\'e algebra do not
have a Galilean interpretation, and will
not be manifest symmetries in the light-cone gauge.

The supercharge $Q_{\cal A}$ is
a $2^{D/2}$ component $SO(D-1,1)$ spinor.
But it decomposes
under the transverse $SO(D-2)$ subgroup
into two (reducible) $2^{(D-2)/2}$ component
spinors playing different roles in the Galilei sub-algebra.
To display this we
choose an appropriate representation for the $\Gamma$
matrices,
convenient for light-cone coordinates.
The $2^{D/2}\times2^{D/2}$ Dirac
gamma matrices satisfy the Clifford algebra
$\{\Gamma^\mu,\Gamma^\nu\}=-2\eta^{\mu\nu}$.
Choose a representation for the
gamma matrices such that $\Gamma^0$ and $\Gamma^{D-1}$ are given by
$$
\Gamma^0=i\pmatrix{0&-I\cr I&0\cr}\qquad\qquad
\Gamma^{D-1}=i\pmatrix{0&0&{\bf 1}&0\cr0&0&0&-{\bf 1}\cr
{\bf 1}&0&0&0\cr0&-{\bf 1}&0&0\cr}\; ,
\(gammas)$$
where $I$ is the $2^{(D-2)/2}$ dimensional identity matrix,
and ${\bf 1}$ is the $2^{(D-4)/2}$ dimensional identity matrix.
This will simplify the superalgebra in light-cone coordinates,
singled out by the spatial
component $D-1$, since $\alpha^{(D-1)}$ is diagonal:
$$\alpha^{(D-1)}\equiv\Gamma^0\Gamma^{D-1}
=\pmatrix{{\bf 1}&0&0&0\cr0&-{\bf 1}&0&0\cr0&0&-{\bf 1}&0\cr
0&0&0&{\bf 1}\cr}\; .\(alpha)$$

The choice of representation for the transverse $\Gamma^k$,
$k=1,\cdots,D-2$ can vary from one dimension to another depending
on whether or not one applies Majorana or Weyl constraints (or both).
Since we only consider even $D$, the Weyl constraint
may always be imposed. If it is, then convenience dictates
a representation for the transverse gamma matrices with the same block
form as $\Gamma^{D-1}$:
$$\Gamma^k=i\pmatrix{0& \gamma^k\cr\gamma^{k}&0\cr},$$
where the $\gamma^k$ are $2^{(D-2)/2}\times2^{(D-2)/2}$
hermitian matrices. In such
a representation
$$\alpha^k=\pmatrix{\gamma^k&0\cr0&-\gamma^k}\; ,$$
and the chirality matrix $\Gamma_{D+1}$ will be diagonal
$$\Gamma_{D+1}=\pmatrix{I&0\cr0&-I\cr}.$$
Imposing the Weyl constraint by
fixing the chirality of the supercharges to be $\pm1$ means
keeping only the first (last) $2^{(D-2)/2}$ components of $Q_{\cal A}$.
On the other hand, if we want the supercharges to be hermitian,
we must choose the $\Gamma^k$ to be imaginary (Majorana).
Only if $D=2$(mod 8) is this possible within the Weyl-friendly
representation just described. The Majorana representation
is also possible for $D=4$(mod 8), but then at least one of
the transverse gammas will not have the block form of
$\Gamma^{D-1}$, so $\Gamma_{D+1}$ won't be diagonal.
For example, in the case $D=4$, a Majorana
representation for the transverse gamma matrices can be taken to
be
$$\eqalign{
\Gamma^1=&i\pmatrix{0&\sigma^1\cr\sigma^1&0\cr}\qquad\qquad
\Gamma^2=i\pmatrix{-I&0\cr0&I\cr}.\cr
}
$$
The Weyl-friendly representation for $D=4$ would retain the same
form for $\Gamma^1$ but replace $\Gamma^2$ by
$$\Gamma^2\rightarrow i\pmatrix{0&\sigma^2\cr\sigma^2&0\cr}.$$

The above representation of the Clifford algebra
helps us display the Galilei properties of the supercharge $Q_{\cal A}$.
This amounts to describing the embedding $SO(D-2)\times SO(1,1)\subset
SO(D-1,1)$ singled out by the light-cone.
Separate the values of ${\cal A}$ into
two groups denoted by dotted and undotted capital Latin spinor indices,
 according to the
eigenvalues of the matrix $\alpha^{D-1}$ \(alpha), the chirality
matrix for $SO(1,1)$:
$$\eqalign{
\alpha^{D-1}_{\dot A\dot B}=& - \delta_{\dot A\dot B}\cr
\alpha^{D-1}_{AB}=& \delta_{AB}\cr
\alpha^{D-1}_{A\dot B}=&\alpha^{D-1}_{\dot AB}=0\; .\cr
}
$$
The dotted and undotted indices each range over
$2^{(D-2)/2}$ values (16 for $D=10$, 2 for $D=4$).
Because the transverse $\balpha$ anti-commute with $\alpha^{D-1}$, it
follows that $\balpha_{\vphantom{\dot A}A\vphantom{\dot B}B}
=\balpha_{\dot A\dot B}=0$.
The spinor supercharge $Q_{\cal A}$
then has dotted components $Q_{\dot A}$, and
undotted components $Q_A$, transforming (reducibly)
as spinors of $SO(D-2)$. The
superalgebra in light-cone coordinates can now be expressed in terms of
these spinors. For later convenience we define $R_{\dot A}\equiv Q_{\dot
A}/\sqrt{2}$.
In terms of the supercharges $Q_A$ and $R_{\dot A}$ the part of the
Super-Galilei algebra involving odd generators is given by
$$\eqalign{
[P^i,Q_A] &= [H,Q_A]=0 \cr
[J^{ij},Q_A] &= - {1\over 2}\Sigma^{ij}_{AB}
  Q^{\phantom{ij}}_{B\phantom{A}} \cr
[K^i,Q_A] &= \vphantom{1\over 2}0 \cr}
\hskip 10pt
\eqalign{
[P^i,R_{\dot A}] &= [H,R_{\dot A}] = 0 \cr
[J^{ij},R_{\dot A}] &= - {1\over 2}\Sigma^{ij}_{{\dot A}{\dot B}}
  R^{\phantom{ij}}_{\dot B\phantom{\dot A}} \cr
[K^i,R_{\dot A}] &= - {i\over 2}\alpha^i_{{\dot A}B}
   Q^{\phantom{i}}_{\vphantom{\dot B}B} \cr}
\hskip 10pt
\eqalign{
\{Q^{\phantom{\dagger}}_A,Q^{\dagger}_B\} &= P^+\delta_{AB}\cr
\{Q^{\phantom{\dagger}}_{\vphantom{\dot A}A},R^{\dagger}_{\dot B}\}
 &= {1\over2}{\bf P}\cdot\balpha_{A\dot B}\cr
\{R^{\phantom{\dagger}}_{\dot A},R^{\dagger}_{\dot B}\}
 &= \half H\delta_{\dot A\dot B}\; .\cr}
\(lcsuperalg)$$
This superalgebra is called $\sgtwo$, where the ``$2$'' stands
for the two supercharges $Q,R$.
In the Weyl friendly representation described above
the spinors $Q_{\vphantom{\dot A}A}$,
$R_{\dot A}$
each decompose into two inequivalent irreducible spinor representations
of $SO(D-2)$, characterized by opposite values of
$\Gamma_{D+1}\alpha_{D-1}$, the chirality matrix for
$SO(D-2)$. To describe this we introduce dotted and undotted
lower case Latin indices according to whether this chirality
matrix has value $-1$ or $+1$, respectively:
$$R_{\dot A}=\pmatrix{R_{\dot a}\cr R_a}\qquad\qquad
Q_A=\pmatrix{Q_a\cr Q_{\dot a}\cr}\; .$$ Then the
$2^{D/2}$ component supercharge $Q_{\cal A}$ breaks up
in our chosen basis as follows:
$$Q_{\cal A}=\pmatrix{Q_a\cr
\sqrt2R_{\dot b}\cr \sqrt2R_c\cr Q_{\dot d}}\; .$$
If the Weyl condition is used to
reduce the spinors, which
means keeping the top (or bottom) two entries, we simply
replace $\alpha^i_{{\dot A}B}\rightarrow\gamma^i_{{\dot a}b}$
(or $-\gamma^i_{{a}\dot b}$) and
$\Sigma^{ij}_{AB}\rightarrow\sigma^{ij}_{ab}
\equiv{-i}[\gamma^i,\gamma^j]_{ab}/2$ (or $\sigma^{ij}_{\dot a\dot b}$)
in \(lcsuperalg).

The Super-Galilei subalgebra of
the Super-Poincar\'e algebra will be relevant
in describing the dynamics of superstring bits.
In fact it will be the full
spacetime symmetry of a field theory of these point-like constituents of
light-cone superstring. The bits are
non-relativistic particles living in
the $(D-2)$ dimensional transverse space, with time given by $x^+$. They
do not know about the longitudinal
direction $x^-$, and consequently there is
no room for the  $M^{-\mu}$ Lorentz generators.
However all information of
the longitudinal direction is not lost.
When bits form into a long polymer,
the conserved bit number operator becomes a candidate
for a discretized $P^+$.
In the limit of infinitely
long polymers, this `$P^+$' is effectively continuous
and the polymers behave as continuous strings moving
in $D$ dimensional space-time, since $x^-$ is conjugate to $P^+$.
With the formation
of infinitely long polymers,
the effective dimension of space is increased by one and, at the
same time, the Galilean invariance is promoted {\bf in
the critical dimension} to a full Poincar\'e
invariance. For the supersymmetric case, it is not immediately
obvious how much of the Poincar\'e superalgebra should be
retained in the superbit dynamics. At first glance, one might
hope to retain the complete superalgebra
displayed in Eq.\(lcsuperalg).
We shall find that this may be too much symmetry for a satisfactory
explanation of string, so we should ask
how much supersymmetry
can be given up while still retaining the full Galilean
symmetry.
It is clear from Eq.\(lcsuperalg) that
one cannot discard the $Q$ supersymmetries without also
discarding the $R$'s. However it {\bf is} consistent
to discard the $R$ supersymmetries while retaining
the $Q$'s. This would correspond
to the Super-Galilei algebra $\sgone$.\[puzalowski,bergmantgal]\
Retaining both dotted and undotted supersymmetries corresponds
to the superalgebra $\sgtwo$.
\chapter{Discrete Superstring in Light-Cone Gauge}
We start with the Green-Schwarz
formulation\[greenssustring,greensbook]\ of
closed superstring theory
in light-cone gauge. The bit model is then motivated
by first constructing a
discretized version of
string on the light-cone.
In the light-cone gauge the world-sheet
reparameterization
invariance
is fixed by choosing $x^+=\tau$ and
choosing $\sigma$ such
that the `+' component of momentum density is constant,
${\cal P}^+=T_0$ with $T_0$ the string rest tension.
\section{II-B}
The light-cone world-sheet variables of
type II-B
superstring theory in $D=10$ space-time dimensions
include, in addition to the coordinates and momenta, the
right- and left-moving Majorana-Weyl spinors  $S^a$
and ${\tilde S^a}$,
transforming in equivalent representations of $SO(8)$, and
obeying anti-commutation relations
$$\eqalign{
\{S^a(\sigma),S^b(\sigma^\prime)\}
=&\delta^{ab}\delta(\sigma-\sigma^\prime)\cr
\{{\tilde S^a}(\sigma),{\tilde S^b}(\sigma^\prime)\}
=&\delta^{ab}\delta(\sigma-\sigma^\prime)\; .\cr}
\(essaessb)
$$
Here the indices refer to the undotted indices of a
fixed chirality ($\Gamma_{11}=+1$) as described in the
previous section, and take the values $1,\ldots,8$.
The light-cone Hamiltonian is given by
$$H=P^-={1\over 2T_0}\int_0^{P^+/T_0}d\sigma
\big[({\cal P}^i)^2 + T_0^2(x^{i\prime})^2
        -iT_0S^aS^{a\prime}
 +iT_0{\tilde S}^a{\tilde S}^{a\prime}\big]\; . \(gshamiltonian)$$
The indices $i,j,k$ are used for the
vector representation of $SO(8)$, and
the indices $a,b,c,d$ and
${\dot a},{\dot b},{\dot c},{\dot d}$ are used for
the two inequivalent spinor representations of $SO(8)$.
The supercharges $Q^a$, ${\tilde Q^a}$,
$R^{\dot a}$, ${\tilde R^{\dot a}}$
generating the $N=2$ supersymmetry
carry both dotted and undotted indices.
The undotted ones are essentially the zero modes of the
spinor variables:
$$Q^a=\sqrt{T_0}\int_0^{P^+/T_0}d\sigma S^a(\sigma)\qquad\qquad
{\tilde Q^a}=\sqrt{T_0}
\int_0^{P^+/T_0}d\sigma {\tilde S^a}(\sigma)\; . \(undottedsusygenc)$$
The dotted components are more complicated bilinears in the
spinor and coordinate variables:
$$\eqalign{
R^{\dot a}=&{1\over2\sqrt{T_0}}\int_0^{P^+/T_0}d\sigma
\gamma^{ib\dot a}S^b(\sigma)({\cal P}^i-T_0x^{\prime i})\cr
{\tilde R^{\dot a}}=&{1\over2\sqrt{T_0}}\int_0^{P^+/T_0}d\sigma
\gamma^{ib\dot a}{\tilde S^b}(\sigma)({\cal P}^i+T_0x^{\prime i})\; .\cr
}
\(dottedsusygenc)
$$

Consider first how the $N=2$ superalgebra is realized.
It is
immediate that all of the $Q$'s and $R$'s  anti-commute with all of the
$\tilde Q$'s and $\tilde R$'s .
It follows from \(essaessb), the canonical commutator of ${\cal P}^i$ and
$x^j$, and periodicity of $x$ in $\sigma$ that
$$\eqalign{
\{Q^a,Q^b\} &= P^+\delta^{ab} \cr
\{Q^a,R^{\dot b}\} &= {1\over 2}{\bf P}\cdot\bgamma^{a{\dot b}}\; , \cr}
\(qalgebra)$$
and similarly for the left-moving supercharges ${\tilde Q}, {\tilde R}$.
To compute the algebra of the $R$ supercharges we will need the following
identities for the $SO(8)$ gamma matrices
$$\eqalign{
\gamma^{i\dot a c}\gamma^{j\dot b c}+(i\leftrightarrow j)=&
2\delta^{ij}\delta^{\dot a\dot b}\qquad\qquad
\gamma^{i\dot a c}\gamma^{j\dot a d}+(i\leftrightarrow j)=
2\delta^{ij}\delta^{cd}\cr
\gamma^{i\dot a c}\gamma^{i\dot b d}+(c\leftrightarrow d)=&
2\delta^{cd}\delta^{\dot a\dot b}\qquad\qquad
\gamma^{i\dot a c}\gamma^{i\dot b c}+(\dot a\leftrightarrow \dot b)=
2\delta^{cd}\delta^{\dot a\dot b}\; .\cr
}\(gammaident)
$$
The top two implement the Clifford algebra, while the bottom two
are Fierz identities which follow from the first two by the special
triality property of $SO(8)$.
We then find that the $R$ supercharges satisfy
$$\eqalign{
\{R^{\dot a},R^{\dot b}\} &= {1\over 4T_0}\int_0^{P^+/T_0}d\sigma
       \big[({\cal P}^i - T_0x^{i\prime})^2 - 2iT_0S^aS^{a\prime}\big]
   = \delta^{{\dot a}{\dot b}}P^-_R \cr
\{\tilde R^{\dot a},\tilde R^{\dot b}\}
&= {1\over 4T_0}\int_0^{P^+/T_0}d\sigma
  \big[({\cal P}^i + T_0x^{i\prime})^2
+ 2iT_0{\tilde S}^a{\tilde S}^{a\prime}\big]
   = \delta^{{\dot a}{\dot b}}P^-_L \; ,\cr}
\(ralgebra)$$
where $P^-_R, P^-_L$ are the right- and left-moving
parts of the light-cone
Hamiltonian respectively, $P^-=P^-_R+P^-_L$.

The above anti-commutators show that
the right- and left-moving supercharges
satisfy independent $N=1$ $\sgtwo$ algebras,
but with {\bf different} Hamiltonians. Thus
an $N=2$ $\sgtwo$ algebra strictly
holds {\bf
only on the subspace of states satisfying the constraint $P^-_R=P^-_L$
($=P^-/2)$.} This is just the $L_0=\tilde L_0$ constraint which
is indeed required in
closed string theory. The first issue we must
settle in discretizing
the world sheet coordinate $\sigma$ is how to treat this constraint.
To do this we note that $L_0-\tilde L_0$ is the generator
of translations in $\sigma$. The states on which it
vanishes are precisely those invariant under this translation.
When $\sigma$ is replaced by a discrete label $k$, the translation
becomes discrete: $k\rightarrow k+1$. Invariance under this
discrete transformation is just a cyclic symmetry requirement
on the string wave function:
$$\Psi(x_1\theta_1,x_{2}\theta_{2},\cdots,x_{M}\theta_{M})
=\Psi(x_2\theta_2,\cdots,x_{M}\theta_{M},x_{1}\theta_{1})\; ,
\(cyclicity)$$
where $\theta_k$ are the Grassmann odd
spinor super-coordinates, defined for
type II-B superstring by
$$\theta^a = {1\over\sqrt 2}\big(S^a - i{\tilde S}^a\big) \; .
\(so8form)$$
In our bit models this symmetry will be an automatic consequence
of the identity of string bits and need not be explicitly imposed.
Since it is a discrete symmetry, it will not have an
infinitesimal interpretation away from the actual continuum
limit, so an analog to the constraint $L_0=\tilde L_0$ will
not exist in the discretized theory, but will
naturally arise in the continuum limit.
{}From this consideration, we see that we need not
{\bf and probably should not} require the full $N=2$ supersymmetry
in our bit model. The $N=1$ supersymmetry generators
$(Q+\tilde Q)/\sqrt2$ and $(R+\tilde R)/\sqrt2$
satisfy the Poinca\'re superalgebra without
constraint, and we
might hope to retain this much supersymmetry
in the discretized theory.

To set up
a model of discrete superstring, we assume that $P^+$ comes
in discrete units $m$, $P^+=Mm$ where $M$ is a large integer
counting the number of bits in a string.
The parameter labeling
points on the string thus becomes discrete
$\sigma\rightarrow
km/T_0$, where $k$ is an integer
taking the values $1,\cdots, M$.
The transverse coordinates are
${\bf x}_k$ corresponding to ${\bf x}(km/T_0)$ and the conjugate
momenta are ${\bf p}_k$ corresponding to $m{\cal P}(km/T_0)/T_0$.
The spinor variables are $S^a_k$ and $\tilde S^a_k$ corresponding to
$\sqrt{m/T_0}S(km/T_0)$ and $\sqrt{m/T_0}{\tilde S}(km/T_0)$
respectively. The non-vanishing (anti)commutators amongst these
discretely labelled variables are:
$$
[x^i_k,p^j_l]=i\delta^{ij}\delta_{kl}\qquad\qquad
\{S^a_k,S^b_l\}=\delta^{ab}\delta_{kl}\qquad\qquad
\{{\tilde S^a_k},{\tilde S^b_l}\}=\delta^{ab}\delta_{kl}\; .
\(discretecoms)$$

The undotted supercharges should obviously be
given by
$$Q^a=\sqrt{m}\sum_{k=1}^MS^a_k\qquad\qquad
{\tilde Q^a}=\sqrt{m}\sum_{k=1}^M{\tilde S^a_k}\; ,\(undottedsusygen) $$
and their algebra is clearly
$$\{Q^a,Q^b\}=mM\delta^{ab}\qquad\qquad\{{\tilde Q^a},
{\tilde Q^b}\}=mM\delta^{ab}\qquad\qquad\{Q^a,{\tilde Q^b}\}=0\; .
\(undottedalgebra)$$
We can also easily guess a discretized form for the $R$'s:
$$\eqalign{
R^{\dot a}=&{1\over2\sqrt{m}}\sum_{k=1}^M
\gamma^{ib\dot a}S^b_k\big(p^i_k-T_0[x^i_{k+1}-x^i_k]\big)\cr
{\tilde R^{\dot a}}=&{1\over2\sqrt{m}}\sum_{k=1}^M
\gamma^{ib\dot a}{\tilde S^b_k}
\big(p^i_k+T_0[x^i_{k+1}-x^i_k]\big)\; .\cr
}\(dottedsusygen)
$$
The anti-commutators of $Q,\tilde Q$ with $R,\tilde R$ are then exactly
of the correct form:
$$\{Q^a,R^{\dot b}\}=\half\bgamma^{a\dot b}\cdot {\bf P}
\qquad\quad\{{\tilde Q^a},
{\tilde R^{\dot b}}\}=\half\bgamma^{a\dot b}\cdot {\bf P}
\qquad\quad\{Q^a,{\tilde R^{\dot b}}\}=
\{{\tilde Q^a},{R}^{\dot b}\}=0\; ,
\(dottedwithnot)$$
where ${\bf P}=\sum_k{\bf p}_k$ is the total transverse
momentum carried by the discretized string.
However $R$ fails to anti-commute with $\tilde R$, breaking
the $N=2$ supersymmetry\foot{Actually it is not hard to modify these
definitions so that $\{R,\tilde R\}=0$:
simply replace $[x^i_{k+1}-x^i_k]$
by $[x^i_{k}-x^i_{k-1}]$ in one ({\bf not both}) of the $R$'s.
But one
would still not get the full $N=2$ algebra because of the constraint
problem mentioned earlier. Even worse, we shall see that the
resolution of the notorious lattice fermion doubling problem, which
is automatic for our choice, would fail for this alternative.}:
$$\{R^{\dot a},{\tilde R^{\dot b}}\}=-{iT_0\over 4m}
                    \sum_{k=1}^M\bgamma^{c\dot a}
                    \cdot\bgamma^{d\dot b}S^c_k\big({\tilde S^d_{k+1}}
                    +{\tilde S^d_{k-1}}-2{\tilde S^d_k}\big)\; .
$$
Using the identities of the $SO(8)$ gamma matrices \(gammaident)
we then derive the rest of the superalgebra,
$$\eqalign{
\{R^{\dot a},{\tilde R^{\dot b}}\}+(\dot a\leftrightarrow \dot b)
                    =&-{iT_0\over 2m}
                    \sum_{k=1}^M\delta^{\dot a\dot b}
                    S^c_k\big({\tilde S^c_{k+1}}
                    +{\tilde S^c_{k-1}}-2{\tilde S^c_k}\big)\cr
\{R^{\dot a},R^{\dot b}\}=&{1\over 4m}
              \sum_{k=1}^M\delta^{\dot a\dot b}
              \Big({\bf p}_k-T_0[\bx_{k+1}-\bx_k]\Big)^2
              -{iT_0\over 2m}\sum_{k=1}^M\delta^{\dot a\dot b}
              S^c_kS^c_{k+1}\cr
\{{\tilde R^{\dot a}},{\tilde R^{\dot b}}\}=&{1\over 4m}
              \sum_{k=1}^M\delta^{\dot a\dot b}
              \Big({\bf p}_k+T_0[\bx_{k+1}-\bx_k]\Big)^2
              +{iT_0\over 2m}\sum_{k=1}^M
              \delta^{\dot a\dot b}{\tilde S^c_k}
              {\tilde S^c_{k+1}}\; .\cr
}
\(dottedalgebra)$$
Although we have lost the full $N=2$ supersymmetry, there remains
an $N=1$ supersymmetry generated by $Q_+=(Q+\tilde Q)/\sqrt2$ and
$R_+=(R+\tilde R)/\sqrt2$. We easily read off the superalgebra
$$\eqalign{
\{Q_+^{a},{Q}_+^{b}\}=&mM\delta^{ab}\qquad\qquad
\{Q_+^{a},R_+^{\dot b}\}=\half\bgamma^{a\dot b}\cdot {\bf P}\cr
\{{R}_+^{\dot a},{R}_+^{\dot b}\}=&{1\over 4m}
              \sum_{k=1}^M\delta^{\dot a\dot b}
      \Big({\bf p}^2_k+T^2_0[{\bf x}_{k+1}
-{\bf x}_k]^2\Big)+{iT_0\over 4m}
      \sum_{k=1}^M\delta^{\dot a\dot b}
{\tilde S^c_k}{\tilde S^c_{k+1}}\cr
&-{iT_0\over 4m}\sum_{k=1}^M\delta^{\dot a\dot b}{S}^c_k{S}^c_{k+1}
      -{iT_0\over 4m}\sum_{k=1}^M\delta^{\dot a\dot b}
      S^c_k\big({\tilde S^c_{k+1}}
+{\tilde S^c_{k-1}}-2{\tilde S^c_k}\big)\; .\cr
}
\(sumalgebra)$$
The last of these equations gives the Hamiltonian,
$$\eqalign{
H=&{1\over 2m}\sum_{k=1}^M\Big[{\bf p}^2_k+T^2_0\big({\bf x}_{k+1}-{\bf
x}_k\big)^2\cr
    &\qquad\qquad -iT_0S^c_kS^c_{k+1} + iT_0{\tilde S^c_k}{\tilde S^c_{k+1}}
     -iT_0 S^c_k\big({\tilde S^c_{k+1}}+{\tilde S^c_{k-1}}-2{\tilde
S^c_k}\big)\Big]\; .\cr}
\(discretegsham)$$
Note that in the continuum limit the last term is
formally sub-dominant to the others since it involves a
second difference. Thus
the Green-Schwarz Hamiltonian \(gshamiltonian)
is regained in the continuum.
This last term, which arises from the nonzero anti-commutator of $R$
with $\tilde R$ is in fact extremely valuable. It breaks world-sheet
chirality in precisely the way needed (\`a la Wilson)
to remove the annoying
fermion doubling problem from the discretized theory. Since the
Hamiltonian is a bilinear form in canonical variables, it is easy
to confirm this through explicit diagonalization of $H$. As always with
quadratic Hamiltonians this is done by finding eigen-operators under
commutation with $H$. Applying this linear operation to each
of the dynamical variables, we find
$$\eqalign{
[H,{\bf x}_k]=&-i{{\bf p}_k\over m}\qquad\qquad
[H,{\bf p}_k]=-i{T_0^2\over m}({\bf x}_{k+1}
+{\bf x}_{k-1}-2{\bf x}_{k})\cr
[H, S^a_k]=&{-iT_0\over 2m}(S^a_{k-1}-S^a_{k+1}
      +2{\tilde S^a_k}-{\tilde S^a_{k+1}}-{\tilde S^a_{k-1}})\cr
[H, {\tilde S^a_k}]=&{+iT_0\over 2m}({\tilde S^a_{k-1}}-
{\tilde S^a_{k+1}}+2S^a_k-S^a_{k+1}-S^a_{k-1}).\cr
}
\(dynamics)$$
To diagonalize these relations we first pass to Fourier modes
$$\eqalign{
{\bf x}_k=&{1\over\sqrt M}
        \sum_{n=0}^{M-1}{\bf \hat x}_n e^{-2\pi ink/M}\qquad\qquad
{\bf p}_k={1\over\sqrt M}
        \sum_{n=0}^{M-1}{\bf \hat p}_n e^{-2\pi ink/M}\cr
{S}^a_k=&{1\over\sqrt M}
        \sum_{n=0}^{M-1}{\hat S}^a_n e^{-2\pi ink/M}\qquad\qquad
{\tilde S^a_k}={1\over\sqrt M}
        \sum_{n=0}^{M-1}{\hat{\tilde S^a_n}} e^{-2\pi ink/M}\; ,\cr
}
\(fourier)$$
with the inverse relations
$$\eqalign{
{\bf\hat x}_n=&{1\over\sqrt M}
        \sum_{k=1}^{M}{\bf x}_k e^{+2\pi ink/M}\qquad\qquad
{\bf\hat p}_n={1\over\sqrt M}
        \sum_{k=1}^{M}{\bf p}_k e^{+2\pi ink/M}\cr
{\hat S}^a_n=&{1\over\sqrt M}
        \sum_{k=1}^{M}{ S}^a_k e^{+2\pi ink/M}\qquad\qquad
{\hat{\tilde S^a_n}}={1\over\sqrt M}
        \sum_{k=1}^{M}{\tilde S^a_k} e^{+2\pi ink/M}.\cr
}
\(inversefourier)$$
One then finds
$$\eqalign{
[H,{\bf\hat x}_n]=&-i{{\bf\hat p}_n\over m}\qquad\qquad
[H,{\bf\hat p}_n]=4i{T_0^2\over m}\sin^2{\pi n\over M}{\bf\hat x}_{n}\cr
[H, {\hat S}^a_n]=&{-iT_0\over 2m}(2i\sin{2\pi n\over M}{\hat S}^a_n
      +4\sin^2{\pi n\over M}{\hat{\tilde S^a_n}})\cr
[H, {\hat{\tilde S^a_n}}]=&{+iT_0\over 2m}
         (2i\sin{2\pi n\over M}{\hat{\tilde S^a_n}}
      +4\sin^2{\pi n\over M}{\hat S}^a_n)\; .\cr
}
\(fourierdynamics)$$
We easily identify the energy lowering operators
$${\bf A}_n={1\over\sqrt{2\omega_n}}
({\bf\hat p}_n-i\omega_n{\bf\hat x}_n)
\qquad\quad
B^a_n=\sin{n\pi\over2M}{\hat S}^a_n
+i\cos{n\pi\over2M}{\hat{\tilde S^a_n}}\; ,
\(lowering)$$
each of which lowers the energy by  the amount $\omega_n/m$
with $\omega_n\equiv 2T_0\sin(n\pi/M)$. Of course
the hermitian conjugates of these operators are energy
raising operators, each of which increases the energy
by the same amount. In the limit
$M\rightarrow\infty$, with $mM$ fixed, finite energy modes occur for
$n$ and $M-n$ finite. These correspond to left- and right-moving
modes respectively, precisely as required for
a continuous closed string.
The excitation energies for these modes are given by
$$ E_n = {2T_0\over m}\sin{n\pi\over M}
={2T_0\over m}\sin{(M-n)\pi\over M} \; ,\(energy)$$
which in the continuum limit with $n$
(or $M-n$) finite approach $2n\pi T_0/P^+$ (or $2(M-n)\pi T_0/P^+$).
Had the $S\tilde S$ coupling term been
absent there would have been additional low energy modes with
$n-M/2$ finite.\foot{The other resolution of the doubling problem
(\`a la Kogut-Susskind)
in which these extra modes are
accepted as part of the physical
spectrum is not satisfactory here because they would include
both integer and half integer modes depending on whether $M$
was even or odd. The half integer modes would ruin the superstring
interpretation.}

The ground state of our discretized string is the one annihilated by
all of the energy lowering operators. The ground state energy
turns out to be exactly zero. (Implying, of course, the absence
of tachyons in the continuum superstring mass spectrum.)
The part of $H=H_{xp}+H_{S\tilde S}$
involving coordinates and momenta, which just describes
a system of harmonic oscillators, applied to the ground state
gives half the sum of all the mode excitation energies:
$$ H_{xp}\ket{G}=\ket{G}{8\over2m}\sum_{n=1}^{M-1}\omega_n
=\ket{G}{8T_0\over m}\sum_{n=1}^{M-1}\sin{n\pi\over M}\; .
\(halfofmodes)$$
The `8' appearing here is just the transverse dimension $D-2$
for ten dimensional space-time.
The part of $H$ involving the spinors gives exactly the
negative of this, with the `8' in this case being the 8 values
of the spinor index $a$, so
$$ H\ket{G}=\ket{G}\left({8T_0\over m}\sum_{n=1}^{M-1}\sin{n\pi\over M}
-{8T_0\over m}\sum_{n=1}^{M-1}\sin{n\pi\over M}\right)=0\; .\(ground)$$
We can summarize the solution of
our discretized superstring model
by quoting the Hamiltonian in terms of raising and lowering
operators:
$$
H={{\bf P}^2\over2mM}+{2T_0\over m}\sum_{n=1}^{M-1}
\sin{n\pi\over M}({\bf A}^\dagger_n\cdot{\bf A}^{\phantom{\dagger}}_n
+B^{a\dagger}_nB^a_n)\; ,
\(fourierham)$$
where ${\bf P}$ is the total momentum.
For completeness we also quote the relation of the dynamical
variables to raising and lowering operators:
$$\eqalign{
{\bf\hat p}_n=&\sqrt{\omega_n\over2}\bigl({\bf A}_n
                                +{\bf A}^\dagger_{M-n}\bigr)
\qquad\qquad\quad\quad~
{\bf\hat x}_n={i\over\sqrt{2\omega_n}}\bigl({\bf A}_n
                                -{\bf A}^\dagger_{M-n}\bigr)\cr
{\hat S}^a_n=&B^a_n\sin{n\pi\over 2M}
+B^{a\dagger}_{M-n}\cos{n\pi\over 2M}
\qquad\quad
{\hat{\tilde S^a_n}}=-i\left(
B^a_n\cos{n\pi\over 2M}-B^{a\dagger}_{M-n}\sin{n\pi\over 2M}\right).\cr
}
\(fouriervars)$$

The discrete II-B superstring model we have presented is the first
step toward a string bit model. Its characteristic feature is
that it has replaced a closed string by a system of $M$
string bits, which are {\bf ordered around a loop}. The
interaction among string bits only exists between nearest
neighbors on this loop. Thus it is not quite a standard
many body system which would allow interactions between all
pairs of particles, and might even include three or more
body interactions. It is very well-known\[thornmosc,thornfock]
how this peculiar pattern
of interactions can arise in a true many body system of
particles described by
$N_c\times N_c$ matrix creation operators in
't Hooft's $N_c\rightarrow\infty$ limit\[thooftlargen]. We shall
turn to this in the next section.

A troubling feature
of the bit-bit interaction from the
string bit point of
view is its long range harmonic form, evident in the Hamiltonian
\(discretegsham). However, it is clear that, as with all
discretizations, the limit
that leads to continuous string should occur for a wide class
of interactions, including ones that are short range. Short
range potentials would of course allow
a discrete string to
dissociate into string bits. All that is necessary to veto
dissociation in the
superstring continuum limit is that the dissociation energy
be of $O(1/m)$ as $M\rightarrow\infty$ with $mM$ fixed.

There are many ways we could introduce a
short range nonharmonic dynamics into our model,
but it is desirable to retain as much of the supersymmetric
structure as possible. One approach is to introduce modifications
into $R,\tilde R$ and then define the Hamiltonian in terms
of these. The simplest possibility is to replace $T_0$ in
\(dottedsusygen)\ by a scalar function
${\cal V}(|{\bf x}_{k+1}-{\bf x}_k|)$.  This has the virtue
of leaving the anti-commutator $\{Q^a_+,R^{\dot b}_+\}$
of the superalgebra undisturbed. We can also allow
a generalization of the spinor structure of the interaction
terms in \(dottedsusygen)\ compatible with $SO(8)$
invariance and the preservation of
$\{Q^a_+,R^{\dot b}_+\}$. For definiteness in this paper we shall forego
such generalizations and restrict to the
following form for $R_+$
$$
R_+^{\dot a}={1\over2\sqrt{2m}}\sum_{k=1}^M \bgamma^{b\dot a}\cdot
\Big[\big(S^b_k+{\tilde S}^b_k\big){\bf p}_k
- \big(S^b_k-{\tilde S}^b_k\big)
\big({\bf x}_{k+1}-{\bf x}_k\big)
{\cal V}
\big(|{\bf x}_{k+1}-{\bf x}_k|\big)\Big].
\(ansatz)$$
Unfortunately,
with ${\cal V}$ not a constant
$\{R_+^{\dot a},R_+^{\dot b}\}$
is no longer proportional to $\delta^{{\dot a}{\dot b}}$,
so that part of the superalgebra will be lost.
In this
situation we propose to define the Hamiltonian by the
positive $SO(8)$ invariant bilinear form
$$H\equiv{1\over4}\sum_{\dot a=1}^8\{R_+^{\dot a},R_+^{\dot a}\}\; .
\(hamgenpot)
$$
This expression automatically commutes with the $Q^a_+$ so
the ${\cal S}_1{\cal G}$ supersymmetry is preserved.
Instead of being the square of a Grassmann odd operator, as would
be a consequence of ${\cal S}_2{\cal G}$ supersymmetry, $H$
has the somewhat weaker property of
being a sum of squares of eight odd operators. By maintaining this
structure we hope to make more likely the recovery of the
full Poincar\'e supersymmetry in the stringy physics. The
structure also naturally guarantees that the energy spectrum
is bounded from below. For the special case ${\cal V}=T_0$,
$H$ reduces to the original
form. Thus we can assert that a satisfactory free superstring
limit will exist provided ${\cal V}$ behaves as a nonzero
constant as far as low energy collective excitations are
concerned.

\section{Not II-B}
Type II-B superstring studied in the
previous section was particularly
neat because of the symmetry between left- and right-moving
waves on
a string. This circumstance allowed a very
appealing resolution of the fermion doubling problem, because
one can form the $SO(8)$ invariant coupling term $S\tilde S$
which raised the energy of the unwanted extra low-lying modes with
mode number near $M/2$. When this left-right symmetry is absent, as in
the type II-A and heterotic superstring theories, another scheme must be
devised to get a satisfactory discretization.

For
type II-A superstring the right-
and left-moving spinors $S,\tilde S$
transform under inequivalent representations
of $SO(8)$. Consequently, the coupling term $S\tilde S$ is not
$SO(8)$ invariant. Therefore one must break the
transverse space rotational symmetry in order to get
rid of the fermion doubling. In fact,
defining canonical spin variables requires a decomposition of the above
spin variables with respect to an
$SU(4)\times U(1)$ subgroup of $SO(8)$,
$$\eqalign{
  \theta^A &= {1\over\sqrt 2}\big(S^A + iS^{A+4}\big)\qquad\qquad
  \pi_A = {1\over\sqrt 2}\big(S^A - iS^{A+4}\big)\; ,\cr}
\(su4form)
$$
and similarly for the left-movers.
The superscript $A=1,\ldots,4$ labels a
$\bf 4$ of $SU(4)$, and the
subscript $A$ labels a $\bar{\bf 4}$.
The decomposition of the representations
is as follows,
$$\eqalign{
  {\bf 8}_s &\rightarrow {\bf 4}_{1/2} + {\bar{\bf 4}}_{-1/2}
\qquad\qquad
  {\bf 8}_c \rightarrow {\bf 4}_{-1/2} + {\bar{\bf 4}}_{1/2} \; ,\cr}
\(decomp)$$
where ${\bf 8}_s,{\bf 8}_c$ are the
two inequivalent spinor representations of
$SO(8)$. (For a detailed discussion see Chapter 11
of \Refer{greensbook}.)
Any coupling between the two kinds of spinor
would have to break either $SU(4)$
or $U(1)$. One can think of the $SU(4)\sim SO(6)$
as the group of rotations in
six ``internal'' dimensions, and the $U(1)$ as the helicity in ordinary
four-dimensional space-time. In this view it is
preferable to preserve the
$U(1)$ symmetry at the discrete level,
even at the cost of breaking the $SU(4)$.

For
heterotic superstring the situation seems even more complicated,
since it has {\bf only} right-moving spinor waves. However as we shall
soon see there may be a more elegant, $SO(8)$ invariant, method to avoid
the fermion doubling problem. This method may also be applied to
type II-A superstring as an alternative to breaking the $SO(8)$ symmetry.

We start by reminding the reader how the doubling problem
arises. Consider the part of the Hamiltonian \(discretegsham)\
involving only the $S$ spinors
$$H_S=-iT_0S^a_kS^a_{k+1}\; ,\(spinham)$$
which is all we would have in the heterotic case where $\tilde S$
is absent. The Fourier modes $\hat S_n$ then satisfy
$$
[H_S,\hat S^a_n]={T_0\over m}\sin{2\pi n\over M}\hat S^a_n
= {2T_0\over m}\sin{\pi n\over M}\cos{\pi n\over M}\hat S^a_n\; .
\(spindynamics)$$
We see that the excitation energies are of $O(1/M)$ not only
for the desired cases of finite $n, M-n$, but also for finite $n-M/2$.
For $n<M/2$ $\hat S^a_n$ raises the energy and is
multiplied in its contribution
to $S^a_k$ by the time dependent phase
$\exp{(+iE_nt-2\pi ink/M)}$ with $E_n>0$.
For $M\rightarrow\infty$ with finite $n$ this corresponds to
a right-moving wave. But in this limit with finite $(M/2)-n$,
the unwanted ``doubled mode'' excitation is a left-moving wave.
Moreover, if $M$ is odd, it acts like a half-integer (anti-periodic)
left-moving mode.
For $M/2<n<M$, $E_n$ is negative (the modes are
energy lowering operators) and those with finite $M-n$
are right-movers for
a continuous closed string
whereas those with finite $n-M/2$ are left-movers.
 Clearly the Kogut-Susskind resolution of the
doubling problem, which is to use the doubled modes as  a part of the
observable physical modes, would wreck the ``heterotic'' nature
of the model:
a continuous closed string would end up with both left-
and right-moving spinor modes. Thus the Wilson alternative which
worked in the II-B case must somehow be used here.

At the moment, the only way we see to do this is to reintroduce
$\tilde S$
as an auxiliary field at the discretized level in such a way
that it resolves the doubling problem but does not
propagate in the continuum limit.
Although we shall not try to develop the
type II-A and heterotic superbit models in this paper, we illustrate how
this might work by examining a spinor model with left-right
asymmetry, described by the Hamiltonian
$$H_{S\tilde S}=-iT_0S^a_kS^a_{k+1}+i\eta T_0\tilde S^a_k\tilde S^a_{k+1}
-i\xi T_0S^a_k\big(\tilde S^a_{k+1}
+\tilde S^a_{k-1}-2\tilde S^a_{k}\big).
\(newspinham)$$
Passing to Fourier modes we have
$$\eqalign{
[H_{S\tilde S}, {\hat S^a_n}]=&{-2iT_0\over m}\sin{n\pi\over M}
(i\cos{\pi n\over M}{\hat S^a_n}
      +\xi\sin{\pi n\over M}{\hat{\tilde S^a_n}})\cr
[H_{S\tilde S}, {\hat{\tilde S^a_n}}]=&{+2iT_0\over m}\sin{n\pi\over M}
         (i\eta\cos{n\pi\over M}{\hat{\tilde S^a_n}}
      +\xi\sin{n\pi\over M}{\hat S^a_n})\; .\cr
}\(newspinmodes)
$$
We find that
$B_n^a=({\hat S^a_n}+\alpha{\hat{\tilde S^a_n}})/\sqrt{1+|\alpha|^2}$
is an energy lowering operator provided
$$\alpha=i\left[{1+\eta\over2\xi}\cot{n\pi\over M}+
\sqrt{1+\left({1+\eta\over2\xi}\right)^2\cot^2{n\pi\over M}}\right]\; ,$$
in which case it lowers the energy by an amount
$$E_n={\omega_n\over m}\left[-{1-\eta\over2}\cos{n\pi\over M}+
\sqrt{\xi^2\sin^2{n\pi\over M}
+\left({1+\eta\over2}\right)^2\cos^2{n\pi\over M}}\right].
\(modeenergy)$$
As long as $\xi\neq0$ and is real and $\eta>0$ there are no
low energy modes other than the ones for finite $n$ and finite $M-n$,
and the doubling problem is avoided. We might
as well simplify matters and take $\xi=(1+\eta)/2$. Then
$$\alpha=i\left[\cot{n\pi\over M}+\csc{n\pi\over M}\right]$$
and
$$E_n={\omega_n\over m}\left[
{1+\eta\over2}-{1-\eta\over2}\cos{n\pi\over M}\right].
\(simpleenergy)$$
The energy lowering operators are then simply
$B^a_n=\sin{n\pi\over 2M}{\hat S^a_n}
+i\cos{n\pi\over 2M}{\hat{\tilde S^a_n}}$.
As $M\rightarrow\infty$, the left-moving modes (finite $n$) have
energy $\eta\omega_n/m$ whereas the right-movers (finite $M-n$)
have energy
$\omega_n/m$, the former a factor of $\eta$ times the latter.
As $\eta\rightarrow\infty$, the left-moving waves gain infinite
energy and would disappear from the spectrum. The discrete theory
could have $\eta$ finite but depend on $m$ in a way that blows
up as $m\rightarrow0$.

Extending this trick to
type II-A is straightforward.
Simply introduce
two additional oppositely moving spinor variables,
with a Hamiltonian similar
to \(newspinham) , except that the new physical spinor is
left-moving and the
new auxliary spinor is right-moving.
A type II-A superstring is thus
constructed as sort of a combination of right-moving and left-moving
heterotic superstrings.
\chapter{Second-Quantized Superstring Bits}
As we saw in the previous section,
discrete light-cone superstring
seems to be made up of non-relativistic interacting
superparticles carrying spin degrees of freedom
and moving in
$(D-2)+1$ dimensional space-time.
If this picture is taken seriously,
a superstring is really a composite object, namely a long
closed polymer of infinitesimal string bits.
Each of these bits
is described by dynamical variables
given by its position
$\bx_k$, its momentum ${\bf p}_k$,  and spin variables
which can be represented in terms of anti-commuting
Grassmann variables $\theta^a_k$,
and their conjugates $\pi^a_k=d/d\theta^a_k$.
The possible states of
a non-interacting (free)
superstring are then given by those of an
$M$-bit polymer, represented by wave functions
$$\Psi({\bf x}_1\theta^a_1,{\bf x}_{2}\theta^a_{2},\cdots,{\bf x}_{M}
\theta^a_{M})
\(firstqwf)
$$
subject to the constraint of cyclic symmetry \(cyclicity).

\section{$1/N_c$ Expansion and Polymers}
According to the Hamiltonian for
a discrete light-cone free superstring
\(discretegsham)\ each bit interacts only with its nearest neighbors.
In order to achieve this nearest neighbor interaction structure in a
second-quantized formulation it is necessary to introduce a ``color''
degree of freedom.
The creation operators
for superstring bits are then $N_c\times N_c$ matrices
transforming in
the adjoint representation of
$U(N_c)$,
$$
\Phi^\dagger({\bf x},\theta)_\alpha^\beta=\sum_{n=0}^{D-2}{1\over n!}
\phi^\dagger_{a_1\cdots a_n}({\bf x})_\alpha^\beta
\theta^{a_1}\cdots\theta^{a_n}\; ,
\(supercreationop)
$$
where the $\phi^\dagger$'s are completely antisymmetric in their
spinor indices
$a_1\cdots a_n$,
and the matrix labels $\alpha,\beta$ run from 1 to $N_c$.
$\phi^\dagger$ creates a boson or fermion according to
whether the number of indices $n$ is even or odd respectively.
The upper limit on $n$ is taken to be $D-2$ because
supersymmetry requires the number of components in
the spinor $\theta^a$ to equal the
number of transverse coordinates. This is of
course possible only for $D=4,6,$ and $10$.
For $D=10$ (or $D=4$) there are all together 256 (or 4)
components of $\phi^\dagger$,
128 (or 2) bosonic
and 128 (or 2) fermionic.
The supercreation operator $\Phi^\dagger$ will
always be Grassmann even and
enjoy
commutation relations. The $\phi^\dagger$'s will of course
satisfy the graded bracket relations,
$$\eqalign{
\big[{\phi}^{\phantom{\dagger}}_{a_1\cdots a_n}({\bf x})_\alpha^\beta &,
{\phi}^\dagger_{b_1\cdots b_m}({\bf y})_\gamma^\delta\big]_\pm \cr
&\equiv
{\phi}^{\phantom{\dagger}}_{a_1\cdots a_n}({\bf x})_\alpha^\beta
{\phi}^\dagger_{b_1\cdots b_m}({\bf y})_\gamma^\delta
 -(-)^{nm}
{\phi}^\dagger_{b_1\cdots b_m}({\bf y})_\gamma^\delta
{\phi}^{\phantom{\dagger}}_{a_1\cdots a_n}({\bf x})_\alpha^\beta\cr
 &= \delta_{mn}\delta_\alpha^\delta
\delta_\gamma^\beta\delta({\bf x}-{\bf y})\sum_P(-)^P\delta_{a_1b_{P_1}}
\cdots\delta_{a_nb_{P_n}}\; .\cr}
\(supercancr)$$
The string bit Fock space is built by
acting on the vacuum state $\ket{0}$
with products of the various creation operators, and consists of states
transforming in various representations of $U(N_c)$. Singlet states are
created by products of traces of matrix products of
the matrix creation operators.
Each trace creates a  closed chain of bits.
We identify the discrete free single superstring wave function
$\Psi$ with the singlet state $\ket{\Psi}$ in
the Fock space of string bits given by
$$\ket{\Psi}=\int\prod_{k=1}^M\big(d^{D-2}x_kd^{D-2}\theta_k\big)
\Tr[\Phi^\dagger({\bf x}_1\theta_1)
\cdots\Phi^\dagger({\bf x}_M\theta_M)]\ket{0}
\Psi({\bf x}_1\theta_1,\cdots,{\bf x}_M\theta_M).
\(stringfock)
$$
Note that once we agree that our state space is the bit Fock
space, {\bf the cyclic symmetry restriction \(cyclicity)\ is
an automatic consequence of the identity of string bits and
the cyclic property of the trace}. Non-interacting multi-string states
would contain a product of several such trace structures.

The world-sheet dynamical variables described in the previous section
are linear differential
operators acting on the single superstring wavefunction $\Psi$.
On our Fock space we seek to represent these operators
as $U(N_c)$ singlets, \ie\ as traces of products of bit
creation and annihilation operators.
To find the bit Fock space representation
of any such dynamical variable $\Omega$, first write down the ket
corresponding to $\Omega\Psi$. Then by an integration by parts transfer
the differential
operator to the creation operators appearing in
the
trace. Finally, one must identify the function of creation and
annihilation operators that reproduces the action of this differential
operator.
Note that once we have a Fock space representation of an operator, it can
act on {\bf any} Fock state, not just singlets. Its action on the single
superstring state \(stringfock)
should however reproduce the action of the
corresponding differential operator on the
superstring wavefunction.

For single body operators
like $Q^a$ and the momentum dependent part of $R^{\dot a}$,
which involve the super-coordinates
of only one bit at a time, the Fock space representation is standard.
Consider for simplicity only a single component matrix creation operator
$a^\dagger(x)_\alpha^\beta$. If we denote
the one body differential operator
$\omega_1$, its Fock space representation will be given by
$$\Omega_1 = \int dx\Tr\big[a^\dagger(x)\omega_1a(x)\big] \; .
\(onebody)$$

For two body operators describing {\bf nearest neighbor}
interactions,
like the coordinate dependent part of $R^{\dot a}$, the identification of
the Fock space representation is not exact. This is because the
second-quantized operators will give interactions between
{\bf all pairs} of
bits. We are therefore led to an approximate treatment using
't Hooft's $1/N_c$ expansion.\[thooftlargen]
To illustrate how this works,\[thornfock]\ consider again the simplified
case of a single component matrix creation operator
$a^\dagger(x)_\alpha^\beta$.
Then the sort of two body operator we will need has the structure
$$\Omega_2={1\over N_c}\int dxdyV(y-x)
\Tr[a^\dagger(x)a^\dagger(y)a(y)a(x)]\; .
\(twobody)$$
Applying this operator to the
singlet Fock state
$\ket{M}=\Tr[a^\dagger(x_1)
\cdots a^\dagger(x_M)]\ket{0}$,
we get after one contraction
\goodbreak
$$\displaylines{
\Omega_2\ket{M}={1\over N_c}\int dy\sum_k V(y-x_k)\hfill\cr
\hfill\times
\Tr\big[ a^\dagger(x_k)a^\dagger(y)a(y)a^\dagger(x_{k+1})\cdots
a^\dagger(x_M)
a^\dagger(x_1)\cdots a^\dagger(x_{k-1})\big]\ket{0}\; .
\cr
}
$$
To continue the evaluation
we note that it matters crucially which creation operator the last
remaining $a(y)$ contracts against.
The contraction with $a^\dagger(x_{k+1})$ produces
a factor of $\sum_\alpha\delta_\alpha^\alpha=N_c$ which cancels the
$1/N_c$ out front. All other contractions
fail to provide a factor of $N_c$.
Thus {\bf in the limit $N_c\rightarrow\infty$}
$$\Omega_2\Tr[a^\dagger(x_1)\cdots a^\dagger(x_M)]\ket{0}
\rightarrow\sum_{k=1}^M V(x_{k+1}-x_k)
\Tr[a^\dagger(x_1)\cdots a^\dagger(x_M)]\ket{0},
\(firstquan)$$
which is precisely the desired nearest neighbor interaction pattern.
The non-nearest neighbor contractions change the trace structure of
the state, giving $1/N_c$ times a state with two traces. Thus $1/N_c$
corrections allow a closed polymer chain to rearrange its
bonds and transform to two closed polymer chains. In the continuous
string limit, this is the origin of string-string interactions.
For more details
and examples, see \Refer{thornfock}.

There is actually some freedom in the choice of the second-quantized
two body operator
\(twobody)
which gives in the limit $N_c\rightarrow\infty$ a
nearest neighbor interaction when acting on singlet states.
One can add to $\Omega_2$  terms with
non-consecutive annihilation operators,
such as
$${1\over N_c}\int dxdyf(x-y):
\Tr\big[a^\dagger(x)a(y)a^\dagger(y)a(x)\big]:\;
.$$
This term can be shown to give
$1/N_c$ times a state with two traces, and
is thus subleading in the limit $N_c\rightarrow\infty$.
Such modifications will alter the general
Fock space properties of $\Omega_2$,
but leave unchanged its action on the singlet states in the limit
$N_c\rightarrow\infty$.
In the next two sections we exploit
these features of $N_c\rightarrow\infty$ to construct second-quantized
expressions for the supercharges and Hamiltonian.

\section{A Superstring Bit Model in 2 +1 Dimensions}
Before developing the $8+1$
dimensional bit model for
real $10$ dimensional superstring, let us first construct
a simpler $2+1$ dimensional model. Long
closed
polymers in this lower dimensional
model will {\bf not} become $4$ dimensional
relativistic strings in the continuum limit,
since the full Poincar\'e algebra
is realized only in $10$ dimensions.
But there are two reasons to study this
model:
\item{1.}It contains all of the features
contained in the higher dimensional
model, but with fewer indices. Thus it
serves as a pedagogical step towards
the higher dimensional model.
\item{2.}We would eventually like to describe strings propagating in $4$
space-time dimensions + $6$ compactified space dimensions. This might be
achieved by such a $2+1$ dimensional bit model with additional internal
degrees of freedom.

Putting aside for a moment the issue of critical dimension, let us assume
that the light-cone Hamiltonian \(gshamiltonian),
or its discrete version
\(discretegsham),
describes a $4$ dimensional type II-B superstring. The
variables $S,\tilde S$ then transform as $2$
dimensional spinor representations
of $SO(2)$. Since $SO(3,1)$ spinors can be
either Majorana or Weyl, but not
both, $S$ and $\tilde S$ are either real two
component spinors, or complex
one component spinors. For simplicity in matching
with the higher dimensional
model we shall use the former. The real 4
dimensional Majorana representation
of $SO(3,1)$ then breaks as follows:
$${\bf 4}\rightarrow {\bf 2} + {\bf 2} \; .$$
The $\bf 2$'s are 2 dimensional {\bf reducible} spinor representations of
$SO(2)$, and  will be labelled by dotted and undotted
upper case Latin letters.
Recall that lower case Latin indices are
reserved for Weyl-restricted spinors,
which are inconsistent with Majorana spinors in 4 dimensions.
The 2 dimensional representations reduce to the two 1 dimensional irreducible
representations
corresponding to spin $\pm1/2$ in the plane.

There are two ways to define
canonical anti-commuting coordinates:
$$\eqalign{
  \theta^A &= {1\over\sqrt 2}\big(S^A - i\tilde S^A\big)
\qquad\qquad
  \pi^A    = {1\over\sqrt 2}\big(S^A + i\tilde S^A\big)\; ,\cr}
\(SOtwo)$$
or
$$\eqalign{
 \theta &= {1\over\sqrt 2}\big(S^1 + iS^2\big)
\qquad\qquad
 \pi    = {1\over\sqrt 2}\big(S^1 - iS^2\big)\; , \cr}
\(Uone)$$
and similarly for the left-movers $\tilde\theta, \tilde\pi$.
The first choice is
analogous to the $SO(8)$ preserving formalism \(so8form),
appropriate for
describing
type II-B superstring. It defines a pair of two component
$SO(2)$ spinors, and is thus termed the ``$SO(2)$ formalism''.
The second choice is
analogous to the $SU(4)\times U(1)$ formalism \(su4form), appropriate
for describing both type II-A and II-B superstring. It defines
two complex Grassmann variables and their canonical conjugates,
and is thus termed the ``$U(1)$ formalism''. Note that since
$SO(2)\sim U(1)$, the two formalisms are equivalent.
This is not so in ten
dimensions, since $SU(4)\times U(1)\subset SO(8)$.

\subsection{\bf $SO(2)$ Formalism}

\noindent From Eq.\(supercreationop) we see
that the superstring bit creation
operator
in the $SO(2)$ formalism is given by
$$\Phi^\dagger = \phi^\dagger + \phi^\dagger_A\theta^A +
                {1\over 2}\phi^\dagger_{AB}\theta^A\theta^B \; ,
\(creation)$$
where the indices $A,B$ run from $1$ to $2$.
Consequently there are two bosonic
and two fermionic degrees of freedom. Written in terms of the canonical
super-coordinates \(SOtwo), the first-quantized
supercharges \(undottedsusygen),
\(dottedsusygen) become
$$\eqalign{
  Q^A &= \sqrt{m\over 2}\sum_{k=1}^M\big(\theta_k^A + \pi_k^A\big) \cr
  \tilde Q^A &
= i\sqrt{m\over 2}\sum_{k=1}^M\big(\theta_k^A - \pi_k^A\big) \cr
  R^{\dot A} &= {1\over 2\sqrt{2m}}\sum_{k=1}^M\alpha^{iB\dot A}
                \big(\theta_k^B + \pi_k^B\big)\big(p_k^i
                - T_0[x^i_{k+1}-x^i_k]\big) \cr
\tilde R^{\dot A} &= {i\over 2\sqrt{2m}}\sum_{k=1}^M\alpha^{iB\dot A}
                \big(\theta_k^B - \pi_k^B\big)\big(p_k^i
                + T_0[x^i_{k+1}-x^i_k]\big)\; , \cr}
\(canonicalsusy)$$
where the relevant matrix elements
of $\alpha^i$, as defined in section 2 are
$$\pmatrix{\alpha^1_{1\dot1}&\alpha^1_{1\dot2}\cr
\alpha^1_{2\dot1}&\alpha^1_{2\dot2}\cr}
= \pmatrix{1&0\cr 0&-1\cr}\qquad\qquad
\pmatrix{\alpha^2_{1\dot1}&\alpha^2_{1\dot2}\cr
\alpha^2_{2\dot1}&\alpha^2_{2\dot2}\cr} = \pmatrix{0&1\cr 1&0\cr}\; .
\(2dalphas)$$
Recall that even though we have two sets of supercharges,
each generating an
independent $N=1$
$\sgtwo$ superalgebra, together
they {\bf do not}
generate an $N=2$ superalgebra since $R$ and $\tilde R$
fail to anti-commute.
The combinations $Q+\tilde Q$ and $R+\tilde R$
satisfy an $N=1$ $\sgtwo$ superalgebra.
Second quantization then follows the
steps described in the previous subsection.
It is simplest to first find the second-quantized
operators associated with
$\theta^A$ and $\pi^A = d/d\theta^A$. These must satisfy the properties
$$\eqalign{
[\Omega_{\theta^A},\Phi^\dagger({\bf x}\theta)]
=&\theta^A\Phi^\dagger({\bf x}\theta)\cr
[\Omega_{\pi^A},\Phi^\dagger({\bf x}\theta)]
=&-{d\over d\theta^A}\Phi^\dagger({\bf x}\theta)\; ,\cr}
\(properties)$$
where the
`$-$' in the second requirement reflects the fact that
a derivative acting on the first-quantized wave function is
transferred to the second-quantized ket through an integration by
parts. It is easy to confirm the following identifications
$$\eqalign{
\Omega_{\theta^A}=&\int d\bx
\Tr\big[\phi^\dagger\phi^{\phantom{\dagger}}_A -
           \phi^\dagger_{A_1}\phi^{\phantom{\dagger}}_{AA_1}\big] \cr
\Omega_{\pi^A}=&\int d\bx
\Tr\big[\phi^\dagger_A\phi^{\phantom{\dagger}} -
                \phi^\dagger_{AA_1}\phi^{\phantom{\dagger}}_{A_1}\big]
=\Omega_{\theta^A}^\dagger \; .\cr
}
\(Omegas)$$
To avoid confusion we will denote the Fock space representations of
the supercharges $Q$
and $R$ by the script letters $\cal Q$ and $\cal R$.
{}From \(Omegas) it immediately follows that the Fock space
representation of the
$Q$ supercharges is
given by
$$\eqalign{
 {\cal Q}^A &= \sqrt{m\over 2}
\big(\Omega_{\theta^A}+\Omega_{\pi^A}\big) \cr
            &= \sqrt{m\over 2}\int
d\bx\Tr\big[\phi^\dagger\phi^{\phantom{\dagger}}_A
         -  \phi^\dagger_{A_1}\phi^{\phantom{\dagger}}_{AA_1} + {\rm
h.c.}\big]\cr
 {\tilde{\cal Q}}^A &= i\sqrt{m\over
2}\big(\Omega_{\theta^A}-\Omega_{\pi^A}\big) \cr
               &= i\sqrt{m\over 2}
          \int d\bx\Tr\big[\phi^\dagger\phi^{\phantom{\dagger}}_A
         -  \phi^\dagger_{A_1}\phi^{\phantom{\dagger}}_{AA_1} - {\rm
h.c.}\big]\; .\cr}
\(secondquant)$$
These second-quantized supercharges
satisfy the same $\sgone$ algebra as the
first-quantized ones \(undottedalgebra), with the
understanding that the bit number
$M$ is replaced by the usual second-quantized number operator:
$$M\rightarrow\int d\bx\Tr\rho(\bx) \; ,\(numberop)$$
where $\rho_\alpha^\beta\equiv [\phi^\dagger\phi
  +\phi_A^\dagger\phi^{\phantom{\dagger}}_A
+{1\over2}\phi_{AB}^\dagger\phi_{AB}^{\phantom{\dagger}}]_\alpha^\beta$.
This is an automatic feature of second-quantized one-body operators,
but it can also be confirmed directly
from the definitions and \(supercancr).

The $R$ supercharges contain both one body and two body operators. It is
therefore convenient to separate their
Fock space representations accordingly,
$$\eqalign{
 {\cal R}^{\dot A}&= {\cal R}_0^{\dot A} + {\cal R}^{\prime\dot A}
\qquad\qquad
 {\tilde{\cal R}}^{\dot A}= {\tilde{\cal R}}_0^{\dot A}
  + {\tilde{\cal R}}^{\prime\dot A}\; .
 \cr}
\(separateR)$$
The expressions for ${\cal R}_0$ and ${\tilde{\cal R}}_0$
are as simple as
those
for ${\cal Q}$ and ${\tilde{\cal Q}}$,
$$\eqalign{
 {\cal R}_0^{\dot A} &= {-i\over 2\sqrt{2m}}
\int d\bx\alpha^{iB\dot A}\Tr\big[
  \phi^\dagger\partial^i\phi^{\phantom{\dagger}}_B -
     \phi^\dagger_{A_1}\partial^i\phi^{\phantom{\dagger}}_{BA_1} - {\rm
h.c.}\big] \cr
 {\tilde{\cal R}}_0^{\dot A}
&= {1\over 2\sqrt{2m}}\int d\bx\alpha^{iB\dot
A}\Tr\big[
   \phi^\dagger\partial^i\phi^{\phantom{\dagger}}_B -
      \phi^\dagger_{A_1}\partial^i\phi^{\phantom{\dagger}}_{BA_1}
+ {\rm
h.c.}\big]\; .
\cr}\(freeR)$$
The Fock space representations
of the two body operators are less obvious,
especially considering the ambiguity alluded to earlier.
The simplest choice that succeeds
in reproducing the first-quantized free superstring results in the limit
$N_c\rightarrow\infty$ is given by,
$$\eqalign{
{\cal R}^{\prime\dot A} &= {-T_0\over 2N_c\sqrt{2m}}\int d\bx\,d\by\
 \balpha^{B\dot A}\cdot(\by-\bx)\cr
&\qquad\qquad\times\Tr\big[
  \phi^\dagger(\bx)\rho(\by)\phi^{\phantom{\dagger}}_B(\bx) -
     \phi^\dagger_{A_1}(\bx)\rho(\by)
\phi^{\phantom{\dagger}}_{BA_1}(\bx)
     + {\rm h.c.}\big] \crr
{\tilde{\cal R}}^{\prime\dot A}
&= {iT_0\over 2N_c\sqrt{2m}}\int d\bx\,d\by\
 \balpha^{B\dot A}\cdot(\by-\bx)\cr
&\qquad\qquad\times\Tr\big[
  \phi^\dagger(\bx)\rho(\by)\phi^{\phantom{\dagger}}_B(\bx) -
     \phi^\dagger_{A_1}(\bx)\rho(\by)
\phi^{\phantom{\dagger}}_{BA_1}(\bx)
     - {\rm h.c.}\big] \; .\cr}
\(intR)$$
The $\cal R$ supercharges then satisfy
the following algebra with the $\cal Q$
supercharges,
$$\eqalign{
 \{{\cal Q}^A,\tilde{\cal R}^{\dot B}\}
&=\{\tilde{\cal Q}^A,{\cal R}^{\dot B}\}
   = 0 \cr
 \{{\cal Q}^A,{\cal R}^{\dot B}\}
&= {1\over 2}\balpha^{A\dot B}\cdot{\bf P}
     + {T_0\over 2N_c}\int d\bx d\by\balpha^{A\dot B}\cdot(\bx-\by)
       :\Tr\big[\sigma(\bx)\rho(\by)\big]: \cr
 \{\tilde{\cal Q}^A,\tilde{\cal R}^{\dot B}\}
&= {1\over 2}\balpha^{A\dot B}
    \cdot{\bf P}
     - {T_0\over 2N_c}\int d\bx d\by\balpha^{A\dot B}\cdot(\bx-\by)
       :\Tr\big[\sigma(\bx)\rho(\by)\big]: \; ,\cr}
\(QwithR)$$
where
$\sigma_\alpha^\beta\equiv:[\phi\phi^\dagger
  - \phi_A^{\phantom{\dagger}}\phi_A^\dagger
 + {1\over 2}\phi_{AB}^{\phantom{\dagger}}
 \phi_{AB}^\dagger]_\alpha^\beta :$.
The integral term on
the right
hand side of the last two anti-commutators signifies a breakdown of the
left- and right-moving $N=1$ $\sgtwo$ algebras.
We expect that acting on a single string
state in the limit $N_c\rightarrow\infty$ this term will vanish,
in order to
reproduce the correct first-quantized anti-commutators \(dottedwithnot).
It is not immediately obvious from the color
structure of the term that this
would be so, so we shall verify it explicitly:
$$\eqalign{
 \int d\bx d\by (\bx-\by)\Tr
:\sigma(\bx)\rho(\by):
\ket{\Psi} &=
  \int d\bx \bx \int d\by
\Tr:[\sigma(\bx)\rho(\by)-\sigma(\by)\rho(\bx)
]:\ket{\Psi} \cr
  &\sim \int d\bx \bx\Tr\int d\by
[\sigma(\bx)\rho(\by) -
                        \rho(\bx)\sigma(\by)]\ket{\Psi} \cr
  &= \int d\bx \bx\Tr\int d\by
[\sigma(\bx)\sigma(\by) -
                          \rho(\bx)\rho(\by)]\ket{\Psi} \cr
  &\sim \int d\bx \bx\Tr
[\sigma(\bx)-\rho(\bx)]
\ket{\Psi} \cr
 &=0
}\(verify)$$
The second line follows for $N_c\rightarrow\infty$ since we have simply
discarded sub-dominant terms.
The equality in the third line follows from the
fact the
$U(N_c)$ charges given by
$$G_\alpha^\beta = \int d\bx
\big[\sigma(\bx)-\rho(\bx)\big]_\alpha^\beta \; ,
\(UNcharges)$$
annihilate all singlet states. The fourth line again follows for
$N_c\rightarrow\infty$ by discarding sub-dominant terms arising from the
contractions. The last line follows from the equality of the traces of
$\sigma$ and $\rho$.

It is immediate from \(QwithR) that the left+right
combinations ${\cal Q}_+ = ({\cal Q}+\tilde{\cal Q})/\sqrt{2}$,
${\cal R}_+ = ({\cal R}+\tilde{\cal R})/\sqrt{2}$ satisfy the correct
anti-commutation relation,
$$\{{\cal Q}^A_+,{\cal R}^{\dot B}_+\}
= {1\over 2}\balpha^{A\dot B}\cdot{\bf
P}\; ,
\(Q+withR+)$$
suggesting the possibility that $N=1$ $\sgtwo$
survives second quantization.
Recall from Eq.\(sumalgebra) that
it was an exact symmetry of the discrete
superstring model, or equivalently of the first-quantized superstring bit
model. In order for this much
supersymmetry
to survive second quantization
the anti-commutator of  ${\cal R}_+^{\dot A}$ with itself must have the
standard form. This computation yields,
$$\{{\cal R}^{\dot A}_+,{\cal R}^{\dot B}_+\} =
  \{{\cal R}^{\dot A}_{0+},{\cal R}^{\dot B}_{0+}\} +
  \{{\cal R}^{\dot A}_{0+},{\cal R}^{\prime\dot B}_{+}\} +
  \{{\cal R}^{\prime\dot A}_{+},{\cal R}^{\dot B}_{0+}\} +
  \{{\cal R}^{\prime\dot A}_{+},{\cal R}^{\prime\dot B}_{+}\}  \; ,
\(RwithR)$$
where the various terms are given by,
$$\eqalign{
\{{\cal R}^{\dot A}_{0+},{\cal R}^{\dot B}_{0+}\}
&={\delta^{\dot A\dot B}\over
4m}
  \int d\bx\Tr\big[|\nabla\phi|^2 + |\nabla\phi_A|^2
   + {1\over 2}|\nabla\phi_{AB}|^2\big] \cr
\{{\cal R}^{\dot A}_{0+},{\cal R}^{\prime\dot B}_{+}\} &+
          (\dot A\leftrightarrow \dot B)
   = {\delta^{\dot A\dot B}T_0\over 4mN_c}\int d\bx d\by \cr
 & \times\Tr\Big[
    \phi^\dagger(\bx)\rho(\by)\phi(\bx) +
    \phi_A^\dagger(\bx)\rho(\by)\phi_A^{\phantom\dagger}(\bx) -
    {1\over 2}\phi_{AB}^\dagger(\bx)\rho(\by)
              \phi_{AB}^{\phantom\dagger}(\bx) \cr
 &\qquad + i\phi^\dagger(\bx)\phi^\dagger(\by)
              \phi^{\phantom\dagger}_A(\by)\phi^{\phantom\dagger}_A(\bx)
         + i\phi^\dagger_A(\bx)\phi^\dagger(\by)
              \phi^{\phantom\dagger}_B(\by)\phi^{\phantom\dagger}_{BA}(\bx)\cr
 &\qquad - i\phi^\dagger(\bx)\phi^\dagger_A(\by)
              \phi^{\phantom\dagger}_{BA}(\by)\phi^{\phantom\dagger}_B(\bx)
         - i\phi^\dagger_A(\bx)\phi^\dagger_B(\by)
              \phi^{\phantom\dagger}_{CB}(\by)\phi^{\phantom\dagger}_{CA}(\bx)
\cr
 &\qquad + \phi^\dagger_A(\bx)\phi^\dagger(\by)
              \phi^{\phantom\dagger}_A(\by)\phi(\bx)
         + \phi^\dagger_{AB}(\bx)\phi^\dagger(\by)
              \phi^{\phantom\dagger}_A(\by)\phi^{\phantom\dagger}_B(\bx) \cr
 &\qquad - \phi^\dagger_A(\bx)\phi^\dagger_B(\by)
              \phi^{\phantom\dagger}_{AB}(\by)\phi(\bx)
         - \phi^\dagger_{AB}(\bx)\phi^\dagger_C(\by)
              \phi^{\phantom\dagger}_{AC}(\by)
\phi^{\phantom\dagger}_B(\bx) \cr
 &\qquad + {\rm h.c.} \Big] \cr
\{{\cal R}^{\prime\dot A}_{+},{\cal R}^{\prime\dot B}_{+}\} &=
 {\delta^{\dot A\dot B}T_0^2\over 4mN_c^2}
\int d\bx d\by d\bz (\by-\bx)\cdot
 (\bz-\bx)\cr
 & \qquad\times\sum_{n=0}^2{1\over n!}
\Tr\big[\phi^\dagger_{A_1\cdots A_n}(\bx)
 \rho(\by)\rho(\bz)\phi^{\phantom\dagger}_{A_1\cdots A_n}(\bx)\big] \cr
 & + {T_0^2\over 8mN_c^2}\int d\bx d\by d\bz\big[
     \alpha^{iA\dot A}\alpha^{jB\dot B}
   - (i\leftrightarrow j)\big](y-x)^i(z-x)^j\cr
 &
\qquad\times\Tr\big[\phi^\dagger_B(\bx)\rho(\bz)
\rho(\by)\phi^{\phantom\dagger}_A
(\bx)
   - i\phi^\dagger(\bx)\rho(\by)\rho(\bz)\phi^{\phantom\dagger}_{AB}(\bx)
   + {\rm h.c.}\big]\; .
\cr}$$
Since there is a term in the last equation which is not proportional to
$\delta^{\dot A\dot B}$, the $\sgtwo$ algebra is not realized with the
second-quantized supercharges. The term by which it fails
gives rise to sub-leading
contributions when acting on single superstring states in the limit
$N_c\rightarrow\infty$. Consequently the first-quantized supercharges
satisfy an $N=1$ $\sgtwo$ algebra as expected.

As sketched in the previous section, even though we do not have the full
$\sgtwo$ superalgebra from which the Hamiltonian is evident,
we can still define a Hamiltonian in the following way
$$H = \sum_{\dot A=1}^2\{{\cal R}^{\dot A}_+,{\cal R}^{\dot A}_+\} \;
.\(bitham)$$
Due to Eq.\(Q+withR+) this Hamiltonian possesses an $N=1$ $\sgone$
supersymmetry. It is in fact {\bf one} Fock space representation of
the first-quantized Hamiltonian \(discretegsham). As we stated earlier,
the Fock space representation of $R^{\dot A}_+$, and therefore of the
Hamiltonian, is determined only up to terms that give rise to subdominant
contributions when acting on the single superstring state $\ket{\Psi}$ in
the limit $N_c\rightarrow\infty$.
One can then try to add such two body terms
to ${\cal R}^{\prime\dot A}_+$ in the hope
of closing the $\sgtwo$ algebra
correctly,
and ending up with a bit theory possessing the full $N=1$ $\sgtwo$
supersymmetry.
Such extra terms would also change the
structure of interactions among
different strings, which appear as subleading terms in the $1/N_c$
expansion. From the point of view of
critical superstring ($D=10$) these
terms may be necessary to get the correct
superstring scattering amplitudes
in the continuum limit of the bit model.
\section{A String Bit Model for Type II-B Superstring}
Now that we've warmed up with a $2+1$ dimensional
supersymmetric bit model,
let's construct an $8+1$ dimensional bit model for
10 dimensional type II-B superstring.
We shall specify the bit dynamics for
type II-B superstring
by working out the second-quantized versions of the
supersymmetry generators and Hamiltonian.
For
II-B  discrete superstring,
the relation of the spinors $S,\tilde S$ to the Grassmann variables
$\theta,\pi=d/d\theta$ maintains $SO(8)$
covariance:
$$
S^a_k={1\over\sqrt2}\left(\theta^a_k
+\pi^a_k\right)\qquad\qquad
{\tilde S^a_k}={i\over\sqrt2}
\left(\theta^a_k-\pi^a_k\right).
\(S-theta)$$
Let us first give the second-quantized
versions
of the undotted supercharges
$Q^a$, $\tilde Q^a$, which
are examples of one body operators.  As with the $2+1$ dimensional case
we first obtain
$$\eqalign{
\Omega_{\theta^a}=&\int d\bx\sum_{n=0}^7{(-)^n\over n!}
\Tr\phi^\dagger_{a_1\cdots a_n}
\phi^{\phantom{\dagger}}_{aa_1\cdots a_n}\cr
\Omega_{\pi^a}=&\int d\bx\sum_{n=0}^7{(-)^n\over n!}
\Tr\phi^\dagger_{aa_1\cdots a_n}\phi^{\phantom{\dagger}}_{a_1\cdots a_n}
=\Omega_{\theta^a}^\dagger\; ,\cr
}
\(Omegas2)$$
from which we find
$$\eqalign{
{\cal Q}^a=&\sqrt{m\over2}\int d\bx\sum_{n=0}^7{(-)^n\over n!}\Tr\big[
      \phi^\dagger_{a_1\cdots a_n}
\phi^{\phantom{\dagger}}_{aa_1\cdots a_n}
      +{\rm h.c.}\big]\cr
{\tilde {\cal Q}^a}=&i\sqrt{m\over2}
\int d\bx\sum_{n=0}^7{(-)^n\over n!}\Tr\big[
      \phi^\dagger_{a_1\cdots a_n}
\phi^{\phantom{\dagger}}_{aa_1\cdots a_n}
      -{\rm h.c.}\big]\cr
{\cal Q}^a_+=&{1\over\sqrt2}
({\cal Q}^a +{\tilde{\cal Q}}^a)\cr
=&\sqrt{m\over2}\int d\bx\sum_{n=0}^7{(-)^n\over n!}
\Tr[e^{i\pi/4}\phi^\dagger_{a_1\cdots a_n}
\phi^{\phantom{\dagger}}_{aa_1\cdots a_n}
+e^{-i\pi/4}\phi^\dagger_{aa_1\cdots a_n}
\phi^{\phantom{\dagger}}_{a_1\cdots a_n}]\; .
\cr
}
\(2ndqplus)
$$
It is again straightforward to
verify that these satisfy the $\sgone$ algebra
of their first-quantized counterparts.
The number operator is again given by
$M=\int d\bx \Tr\rho(\bx)$, where
the bit density matrix in $8+1$
dimensions is given by
$$
\rho(\bx)_\alpha^\beta = \sum_{n=0}^8{1\over n!}
\big[\phi^\dagger_{a_1\cdots a_n}(\bx)
\phi^{\phantom{\dagger}}_{a_1\cdots
a_n}(\bx)
\big]_\alpha^\beta \; .
\(density)$$

Next we turn to the $R$ supercharges which involve two body operators.
As in the $2+1$ dimensional case we only present
the simplest second-quantized candidates which, in the
limit $N_c\rightarrow\infty$, produce on single polymer states
both $R$ and $\tilde R$ given in \(dottedsusygen).
We generalize slightly, replacing $T_0$ by a general
scalar potential ${\cal V}(|{\bf x}_{k+1}-{\bf x}_k|)$.
Writing ${\cal R}={\cal R}_0+{\cal R}^\prime$ and
$\tilde {\cal R}=\tilde {\cal R}_0+\tilde {\cal R}^\prime$,
where the subscript $0$ denotes the one body term and
prime denotes the two body term, we end up with
$$\eqalign{
{\cal R}_0^{\dot a}=&{-i\over2\sqrt{2m}}\int d\bx\gamma^{ib\dot a}
         \sum_{n=0}^7{(-)^n\over n!}
 \Tr\big[\phi^\dagger_{a_1\cdots
a_n}\partial^i\phi^{\phantom{\dagger}}_{ba_1\cdots a_n}
 - {\rm h.c.}\big]\cr
\tilde {\cal R}_0^{\dot a}=&{1\over2\sqrt{2m}}\int d\bx\gamma^{ib\dot a}
         \sum_{n=0}^7{(-)^n\over n!}
\Tr\big[\phi^\dagger_{a_1\cdots a_n}\partial^i
\phi^{\phantom{\dagger}}_{ba_1\cdots a_n} +{\rm h.c.}\big]\cr
{\cal R}^{\prime \dot a}
 =&{-1\over 2N_c\sqrt{2m}}\int d\bx d\by\sum_{n=0}^7
 {(-)^n\over n!}({\bf y}-{\bf x})\cdot\bgamma^{b\dot a}{\cal V}
 (|{\bf y}-{\bf x}|)\cr
 & \qquad\qquad\qquad\times\Tr\big[\phi^\dagger_{a_1\cdots a_n}({\bf x})
 \rho(\by)
 \phi^{\phantom\dagger}_{ba_1\cdots a_n}({\bf x}) + {\rm h.c.}\big]\cr
\tilde {\cal R}^{\prime \dot a}
 =&{i\over 2N_c\sqrt{2m}}\int d\bx d\by\sum_{n=0}^7
 {(-)^n\over n!}({\bf y}-{\bf x})\cdot\bgamma^{b\dot a}{\cal V}
 (|{\bf y}-{\bf x}|)\cr
 &\qquad\qquad\qquad\times\Tr\big[\phi^\dagger_{a_1\cdots a_n}({\bf x})
 \rho(\by)
 \phi^{\phantom\dagger}_{ba_1\cdots a_n}({\bf x})
- {\rm h.c.}\big]\; .\cr
}
\(2ndrdots)
$$
Other terms
with non-consecutive creation operators and
with more general spinor structure have not been displayed
here, but we expect such terms are needed to get the
superstring interactions
right.
These supercharges again fail to satisfy the $\sgtwo$ algebra.
It is still conceivable that
with more complicated color routings and spinor
structures
the full $\sgtwo$ supersymmetry can be restored. But this may not
be possible, and we don't think it
should necessarily be required at the level of string bits,
which should generically exhibit less symmetry than the
continuum.
At the first-quantized level
(equivalent to the second-quantized theory at $N_c\rightarrow\infty$)
the full $S_2{\cal G}$ superalgebra for ${Q}_+$ and
$R_+$ was only present for a constant ${\cal V}=T_0$.
For non-constant
${\cal V}$ but unchanged spinor structure, we only
had the $S_1{\cal G}$ algebra generated by ${Q}_+$. For the
second-quantized theory at finite $N_c$
our simplest ansatz for ${\cal R}^\prime_+$ fails to close the
$\sgtwo$ superalgebra because $\{{\cal R}^{\prime\dot a}_+,
{\cal R}^{\prime\dot b}_+\}$ is not proportional to
$\delta^{\dot a\dot b}$.\foot{For ${\cal V}=T_0$, the cross
terms $\{{\cal R}_{0+}^{\dot a},
{\cal R}^{\prime\dot b}_+\}+(\dot a\leftrightarrow\dot b)\propto
\delta^{\dot a\dot b}$.
This is not surprising because this operator has
a color structure that can survive the limit $N_c\rightarrow\infty$,
and we already know from the first-quantized
theory that $\sgtwo$ holds in that limit when
${\cal V}=T_0$.}
The offending contributions,
however, have a color structure which is sub-dominant as
$N_c\rightarrow\infty$.
The
supersymmetry generated by the ${\cal Q}_+$'s remains a symmetry
at finite $N_c$ for any ${\cal V}$ if the
Hamiltonian commutes with ${\cal Q}_+^a$, and we shall insist that
at least $\sgone$ be an exact
symmetry of the string bit dynamics. This will
automatically hold
if we define the Hamiltonian $H$ for the second-quantized
theory by
\(hamgenpot)\ with second-quantized operators
${\cal R}_+$ substituted for
the first-quantized $R_+$.
This is because we not only have the $\sgone$ superalgebra
$$\{{\cal Q}^a_+,{\cal Q}^b_+\}
=\delta_{ab}m\int d\bx\sum_{n=0}^8{1\over n!}
\Tr\phi^\dagger_{a_1\cdots a_n}\phi^{\phantom{\dagger}}_{a_1\cdots a_n}
\equiv\delta_{ab}mM,
\(Q+algebra)$$
but we also require the $\sgtwo$ anti-commutators between ${\cal Q}_+$ and
${\cal R}_+$:
$$\{{\cal Q}^b_+,{\cal R}
^{\dot a}_+\}=\half\bgamma^{b\dot a}\cdot
\int d\bx\sum_{n=0}^8{1\over n!}
\Tr\phi^\dagger_{a_1\cdots a_n}(-i{\bf \nabla})
\phi^{\phantom{\dagger}}_{a_1\cdots a_n}
=\half\bgamma^{b\dot a}\cdot{\bf P},
\(Q+R+algebra)$$
where ${\bf P}$ is the total momentum of the multi-bit
system. It then follows from our definition of $H$
that $[H,{\cal Q}_+^a]=0$, since all the ${\cal R}$'s
are translationally
invariant, and so commute with ${\bf P}$.

We have now presented the ingredients of our proposed
string bit model for
type II-B superstring. To summarize
our results, we recall the steps in the construction of the complete
Hamiltonian. First construct ${\cal R}_+^{\dot a}={\cal R}_{0+}^{\dot a}
+{\cal R}_+^{\prime\dot a}$ from the expressions
listed in \(2ndrdots)\ or from a generalization of them.
For example, using the displayed expressions we obtain
$$\eqalign{
{\cal R}_{0+}^{\dot a}
 =&{1\over\sqrt2}({\cal R}_{0}^{\dot a}
+{\tilde {\cal R}}_{0}^{\dot a})\cr
 =&{1\over2\sqrt{2m}}\int d\bx\gamma^{ib\dot a}
\sum_{n=0}^7{(-)^n\over n!}
 \Tr\big[e^{-i\pi/4}\phi^\dagger_{a_1\cdots a_n}
 \partial^i\phi^{\phantom{\dagger}}_{ba_1\cdots a_n}
+ {\rm h.c.}\big] \cr
{\cal R}_+^{\prime \dot a}=&{1\over\sqrt2}({\cal R}^{\prime\dot a}
 +{\tilde {\cal R}}^{\prime\dot a})\cr
 =&-{1\over 2N_c\sqrt{2m}}\int d\bx d\by\sum_{n=0}^7
 {(-)^n\over n!}({\bf y}-{\bf x})\cdot\bgamma^{b\dot a}{\cal V}
 (|{\bf y}-{\bf x}|)\cr
 &\qquad\qquad\qquad\times\Tr\big[e^{-i\pi/4}
\phi^\dagger_{a_1\cdots a_n}({\bf x})
 \rho(\by)
 \phi^{\phantom\dagger}_{ba_1\cdots a_n}({\bf x}) + {\rm h.c.}\big] \; .\cr
}
\(2ndrplusdots)
$$
Once the ${\cal R}_+$'s are pinned down, our proposal
for the string bit Hamiltonian will be
$$\eqalign{
H_{\rm II-B}=&\fourth\sum_{\dot a=1}^8
\{{\cal R}_{0+}^{\dot a}+{\cal R}_{+}^{\prime\dot a},
{\cal R}_{0+}^{\dot a}+{\cal R}_{+}^{\prime\dot a}\}\cr
=&{1\over 2m}\int d\bx\sum_{n=0}^8{1\over n!}
\Tr|\nabla\phi_{a_1\cdots a_n}|^2
+\half
\{{\cal R}_{0+}^{\dot a},{\cal R}_{+}^{\prime\dot a}\}
+\fourth\{{\cal R}_{+}^{\prime\dot a},{\cal R}_{+}^{\prime\dot a}\}\;,
\cr}\(proposedbitham)
$$
where we have only written out the free part of the string bit
Hamiltonian explicitly. The interacting terms are to be worked out
using \(2ndrplusdots)\ or its generalization.

The Hamiltonian \(proposedbitham)\ defines the dynamics we
propose for string bits, once we have specified
the structure of ${\cal R}_+^\prime$.
For finite $N_c$ it describes
a perfectly well-defined non-relativistic many-body system.
When studied in the limit $N_c\rightarrow\infty$,
it will, by construction,  describe weakly interacting
long polymers  and the
infinitely long ones will have exactly the properties of
type II-B free superstrings. Interactions among strings will
also be included in \(proposedbitham)\ with strength of order
$1/N_c^2$ for the string-string scattering amplitude.
Unfortunately, the string interactions arising
from the terms displayed in \(2ndrplusdots)\
do not seem to provide the richness of spinor structure required
in the light-cone three-superstring vertex given by
Green, Schwarz , and Brink\[greensb]. The basic structure
of the correct three-string vertex term in the
supercharge is an ``overlap'' integral of
the product of the three string wave functions with an
insertion of a complicated seventh order polynomial
of the world-sheet spinors at the joining point.
Inspection of the $1/N_c$ terms arising from non-nearest neighbor
contractions in the action on a polymer state of the terms displayed
in \(2ndrplusdots)\ confirms the
basic overlap structure. But these terms can provide only a
linear factor of world-sheet spinors at the joining point.
Thus it is clear that terms in ${\cal R}$ with a more complicated
spinor structure will be required.\foot{If
they are restricted to terms with nonconsecutive creation operators,
\eg\ with the trace structure $\Tr:\phi^\dagger\phi\phi^\dagger\phi:$,
they will not affect the properties of free strings.}
 This means that the
principles we have so far imposed on our string
bit models are not quite strong enough to force the correct
dynamics for
interacting superstring theory. In
\Refer{greensb}\ it was shown that requiring the
Poincar\'e superalgebra was sufficient to uniquely determine
the three string vertex. Thus, if we could succeed in devising
a string bit model with the full $\sgtwo$ supersymmetry
at finite $N_c$ {\bf and} a large $N_c$ limit that correctly
describes free superstrings, the correct stringy interactions
would be virtually guaranteed. So far we have examples which
fullfil either of these criteria, but not both. The models
given in this paper are constructed to satisfy the second
criterion, but they fall short of the first.
In \Refer{bergmantgal} we construct a model possessing the full
$\sgtwo$ supersymmetry, but it is unlikely that its
large $N_c$ limit describes free superstrings.
Lacking a satisfactory model
with the full $\sgtwo$ supersymmetry, one should adopt the
renormalization group philosophy and allow {\bf all} interactions
consistent with $\sgone$ symmetry and search for the interesting
cases among all possible continuum limits, one of which
should be the interacting type II-B superstring theory.
The various superstring/bit models
and their supersymmetries are summarized in Table 1.

\vskip1.5cm
\begintable
\multispan{4}\tstrut\hfil Superstring/Bit Models\hfil\crthick
Model | ${\cal V}(\bx)$ | SUSY | Failing (anti)-commutators \crthick
Covariant type II | $-$ | $
N=2$ ${\cal SP}$ | none \cr
Light-Cone type II, $D=10$ | $-$ | '' | none \cr
Light-Cone type II, $D\neq 10$ | $-$ | $N=2$ $\sgtwo$ |
$[M^{-i},M^{-j}]\neq 0$
\cr
Discrete Light-Cone | $-$ | $N=1$ $\sgtwo$ |
$\{R^{\dot a},\tilde R^{\dot b}\}\neq 0$
\cr
$1^{st}$ Quantized Super-Bits | ${\cal V}=T_0$ | '' | '' \cr
'' | ${\cal V}(\bx)$ | $N=1$ $\sgone$ |
$\{R^{\dot a}_+,R^{\dot b}_+\}
\not{\hskip-2pt\propto}
\delta^{\dot a\dot b}$ \cr
$2^{nd}$ Quantized Super-Bits | ${\cal V}=T_0$ | '' |
$\{{\cal Q}^a,{\cal R}^{\dot a}\}$,
$\{\tilde{\cal Q}^a,\tilde{\cal R}^{\dot a}\}$,
$\{{\cal R}^{\dot a}_+,{\cal R}^{\dot b}_+\}$ \cr
'' | ${\cal V}(\bx)$ | '' | '' \endtable
\centerline{{\it Table 1.} Superstring models and their supersymmetry.}
\chapter{Ambiguities and Open Issues}
In this paper we have made a proposal for the extension of string bit
models to
superstring, developing most fully the type II-B
case while leaving
the complete analysis of the not II-B cases for future work.
However we have only made a start on
the task of confirming that the proposal reproduces completely all
aspects of superstring theory. While we can firmly assert that the
$N_c\rightarrow\infty$ limit describes free superstrings adequately, the
$1/N_c$ corrections which determine the interactions among strings have
not yet been well studied. It is transparent from the string
bit compositeness that these interactions will be string breaking/joining
processes with amplitudes proportional
to overlap integrals between initial and final multistring states.
We can anticipate also that, as was the case for
bosonic string,
the interactions can only be fully Poincar\'e invariant in the
critical dimension. However the details, including
any modifications required to produce the correct operator
insertions at the joining
points,  have yet to be worked out. These operator
insertions are also known to entail contact
interactions\[greenseiberg,greensiteklink], which should
of course also be a consequence of our string bit model.
We fully expect that terms must be
added to the second-quantized interacting supercharge ${\cal R}$
which do not contribute
in leading order in the $1/N_c$ expansion. Any monomial such as
$\Tr:\phi^\dagger\phi\phi^\dagger\phi:$,
in which the creation operators are
not consecutive is such a term. All of these
issues need to be carefully examined in future work.

Assuming that either our proposed Hamiltonian or a
suitable modification
of it correctly reproduces the interacting
superstring theory, there is still
the question of uniqueness. Because stringy physics is only a
property of infinitely composite string bit polymers, it is natural to
expect, in accord with ideas of universality, that there are
many microscopic string bit models that yield the same
continuum string theory.  One aspect of this is our expectation
that a wide class of potentials ${\cal V}$ will give identical
stringy physics to the case ${\cal V}=T_0$. The degree of
flexibility in the choice of potential still needs to be pinned
down. In particular, our conjecture that the potential could
even be short range needs to be tested
(at the very least numerically). These are all issues
that can be addressed at the first-quantized non-interacting
(\ie\ $N_c\rightarrow\infty$)
polymer level. But they are also pertinent to the fully
interacting (finite $N_c$) string bit theory.
For example, the ambiguities
mentioned in the previous paragraph may to some extent be
string bit artifacts that can be absorbed into
the definition of a small number of macroscopic
string parameters and have no further effect on the string interactions.

Finally we say a few words about compactification, a subject
we have not yet addressed. A string model of the real world must
of course possess precisely 4 noncompact dimensions, 3 space
plus 1 time. This means that the corresponding string bit
model should have precisely 2 noncompact spatial dimensions.
Accordingly, we must eventually ``compactify'' 6 of the
8 spatial dimensions of our superbit models. One possibility
is, of course, to impose by hand that a 6 dimensional subspace
is some compact space, be it a toroid, orbifold, or Calabi-Yau
manifold.
But the string bit picture allows a more dramatic possibility.
Polymer formation generically promotes finite internal degrees
of freedom on the bit to an effective compact dimension\[gilesmt].
Indeed
the manner in which the world sheet spinor fields $S$, $\tilde S$
emerge from the
256 component string bit multiplet illustrates this point very
nicely. A pair of world sheet fermion fields can always be
``bosonized.'' The resulting boson world sheet field then
enters the string dynamics in just the way a compactified
coordinate would. In particular it would count as part of
the $D$ which is required to be 10 for
superstring. In this
way the string bit model might be properly formulated from
the beginning as a 2+1 dimensional
Super-Galilei invariant theory of string
bits, which carry, in addition to the supermultiplet
spin labels, a finite number of
internal degrees of freedom to play the role of the
6 compactified dimensions. A successful implementation
of this possibility would provide an explicit and concrete
realization of 't Hooft's idea\[thooftbh]\ that the world is a hologram:
That it is fundamentally a system existing in 2 spatial
dimensions, although it gives the appearance of being
3 dimensional.
\medskip
\endpage
\titlestyle{References}
\smallskip
\reflist{}
\bye